\documentclass[prb,aps,nofootinbib,superscriptaddress,twocolumn,floatfix]{revtex4-1}

\usepackage{graphicx}
\graphicspath{{nafigures/}} 
\usepackage{epstopdf}
\usepackage{amsmath}
\usepackage{dcolumn}

\begin{document}

\title{Two and three particles interacting in a one-dimensional trap}
\author{MengXing Na}
\email{mengxing@ualberta.ca \phantom{aaa} Present address: Physics and Astronomy Department, University of British Columbia}
\author{Frank Marsiglio}
\email{fm3@ualberta.ca}
\affiliation{Department of Physics, University of Alberta, Edmonton, AB, Canada T6G 2E1}

\begin{abstract}
We outline a procedure for using matrix mechanics to compute energy eigenvalues and eigenstates for two and three interacting particles in a confining trap, in one dimension.  Such calculations can bridge a gap in the undergraduate physics curriculum between single-particle and many-particle quantum systems, and can also provide a pathway from standard quantum mechanics course material to understanding current research on cold-atom systems.  In particular we illustrate the notion of ``fermionization'' and how it occurs not only for the ground state in the presence of strong repulsive interactions, but also for excited states, in both the strongly attractive and strongly repulsive regimes.
\end{abstract}

\maketitle

\section{introduction}

Nowadays undergraduate physics students are increasingly exposed to research-related activities throughout the course of their studies.  This is often done through summer
research fellowships that expose students to hands-on laboratory or theoretical
work.  Moreover, upper-level lab courses often have
a research flavor, being more open-ended than their introductory counterparts.
An equivalent open-endedness tends not to exist in theoretically oriented courses, although in recent decades the increased use of the computer in lessons and homework assignments has slowly been changing this.\cite{examples} 

Meanwhile, in many fields in physics, a modern theme in research is the effect of interactions among the constituent particles.  Undergraduates in the 21st century are well poised to learn more about such problems, mainly in the context of undergraduate quantum mechanics.

First, one should acknowledge that students are already exposed
to particle-particle interactions---it is just that we tend to quickly disguise that this is the case.  For example, the
hydrogen atom is really a two-particle problem, where we (wisely) adopt center-of-mass and relative coordinates, quickly reducing
this problem to that of a single
particle with reduced mass in the presence of an ``external'' potential.  This change of coordinates is a good
thing, insofar it allows us a complete analytical solution to the problem.\cite{schrodinger26} However, it provides no
guidance to what is to be done as the number of particles increases, and it tends to leave the student with the
impression that further progress is impossible and/or requires approximation methods.\cite{ring80}

The few-body problem was originally most relevant in nuclear physics, as indicated by the books cited above.  However, as
activity in this field has diminished, it has been replaced with increased furor in the field of ``cold atoms,'' where lasers
are used to confine particles.  See some recent reviews in 
Refs.~\onlinecite{bloch08,chin10,cazalilla11,zinner15}, along with more pedagogical expositions for the experiments in 
Ref.~\onlinecite{appleyard07} and for the theory in Ref.~\onlinecite{shea09}. This field has exploded over the past decade, 
with increased interest in so-called optical lattices, along with the ability to tune all the relevant interactions, including fine
details like spin-orbit coupling.\cite{leblanc15} Remarkably, both the traps and the lattices can be manipulated to be three-, two-, or 
one-dimensional.  It is mostly because of the arrival of this new ``playground'' for physicists, where
particles with different statistics can be readily utilized, and crossover phenomena from weak to strong interactions can be tuned through manipulation of their Feshbach resonances,\cite{feshbach00} that this explosion has occurred. 

At the same time, these developments involve scenarios that are increasingly accessible to the classroom.  In particular, we will take advantage of the ability to manufacture systems of interacting particles in any dimension to provide a systematic, textbook-like
account of interacting particles in one-dimension.  As in Ref.~\onlinecite{shea09} we will focus primarily on two particles, but with a careful watch on how generalizations to larger numbers can (in principle) be performed.  We
will adopt the matrix mechanics approach used in Ref.~\onlinecite{marsiglio09} to carry out calculations numerically.

Before getting into specific examples, we start with a general Hamiltonian to deal with any confining potential and various forms of the two-body interaction:
\begin{equation}
\hat{H} =  \sum_i \left\{ -{\hbar^2 \over 2m_0} {d^2 \over dx_i^2} + \hat{V}_{\rm conf}(x_i) \right\} + \sum_{i < j} \hat{V}_{\rm int}(|x_i - x_j|),
\label{eq:gen_ham}
\end{equation}
where the sums are over the particles in the system.  The first term of the Hamiltonian is a simple sum of one-body contributions, including the confining potential, denoted $\hat{V}_{\rm conf}$.  The second term contains the two-body interactions,
which we will take to be solely a function of the distance between any two particles, as indicated.  All particles will be taken to have mass $m_0$, but we will reserve for separate consideration the three cases of distinguishable, boson, and fermion statistics.

In the following section we start with a one-dimensional confining potential that is more familiar to undergraduates: the infinite square well.  We examine the procedure for understanding the behavior of more than one particle in such a well, and then move on to the more experimentally relevant harmonic trap.\cite{remark1} We do this in several ways.  First, the least intimidating (from the perspective of 
a novice) method is to ``embed'' the harmonic potential in an infinite square well, and proceed as in the previous case.  The complications
with respect to the infinite square well case then arise only in the one-particle problem, following Ref.~\onlinecite{marsiglio09}. The second
step is to dispense with the infinite square well altogether and simply use the single-particle eigenstates for the harmonic oscillator
as basis states.  Although we must deal with more complicated functions (Hermite polynomials), this choice actually makes
the problem simpler, and conforms with the methodology used in some of the research literature.
The interaction term requires a straightforward integration, which can be made very efficient (in fact, analytical) through a mathematical trick, which we discuss in an appendix. 

We also point out the phenomenon of ``fermionization,'' a process whereby distinguishable particles behave like
fermions when the interactions become particularly strong.\cite{zurn12} We note this phenomenon in both the energies and in the wave functions,
and for both the strongly repulsive and the strongly attractive regimes.

Finally, removal of the center-of-mass degree of freedom simplifies the problem still further, and we outline this procedure at the end of Sec.~\ref{sec_three}. 
The final section is
devoted to a brief discussion of the three-particle problem; this section serves as a ``launching pad'' for addressing the $N$-particle problem.

\section{interactions in an infinite square well\label{sec_two}}

We begin by considering two interacting particles confined by an infinite square well potential.  Although the infinite square well is not the most realistic confining potential, it has the advantage of being familiar to students.

The simplest form of two-particle interaction is the contact potential, that is, a Dirac $\delta$-function.  While this form of the interaction is best for
straightforward evaluation of the required matrix elements, it is also the poorest for convergence as a function of the number of
basis states.  This is because $\delta$-function interactions tend to give rise to ``cuspiness'' in the wave function.  As was demonstrated in Ref.~\onlinecite{marsiglio09}, this difficulty also occurs for the case of a single particle interacting with a $\delta$-function potential.\cite{marsiglio09a}

\subsection{Review of one-particle results}

We review one-particle results (without interactions) for the purpose of establishing notation.\cite{marsiglio09}
The infinite square well (isw) potential of width $a$ for a particle at position $x_i$ is defined as
\begin{equation}
V_\textrm{isw}=\begin{cases}
		0 & 0\leq x_i \leq a, \\
		\infty & \text{otherwise.}
		\end{cases}
\label{Eq: ISW potential}
\end{equation}
The single-particle Hamiltonian is then the sum of the kinetic term and the confining potential term as defined in Eq.~(\ref{eq:gen_ham}), so for a collection of noninteracting particles the Hamiltonian is
\begin{equation}
\hat{H}_0=\sum_i \left\{ \frac{-\hbar^2}{2m_0}\frac{d^2}{dx_i^2}+V_\textrm{isw}(x_i) \right\}.
\label{Eq: ISW Hamiltonian}
\end{equation}
The well-known single-particle eigenstates and eigenvalues for this problem are
\begin{equation}
\phi_n(x_i)=\begin{cases}
		\sqrt{\frac{2}{a}}\sin\big(\frac{n\pi x_i}{a}\big) & 0\leq x_i\leq a,\\
		0 & \text{otherwise,}
		\end{cases} 
\label{Eq: ISW basis}
\end{equation}
and 
\begin{equation}
E_n=\frac{n^2\pi^2\hbar^2}{2m_0a^2}=n^2E_1,
\label{Eq: ISW energies}
\end{equation}
with the quantum number $n=1,2,3,\ldots\,$, and $E_1~\equiv~\pi^2 \hbar^2/(2m_0 a^2)$.

\subsection{More than one particle}

The non-interacting Hamiltonian for two or more particles is simply the one-particle Hamiltonian, Eq.~(\ref{Eq: ISW Hamiltonian}).
Solutions are then built out of the basis consisting of product states of the one-particle basis states of the infinite square
well, Eq.~(\ref{Eq: ISW basis}). 
This means quantum labels begin to proliferate, and also depend on the statistics of the particles.  In principle there are three cases:
distinguishable, fermion, and boson. For example, for two particles, the product wave function in the distinguishable case is
\begin{equation}
\psi_{n_1,n_2}=\phi_{n_1}(x_1)\phi_{n_2}(x_2),
\label{Eq: two particle basis}
\end{equation}
where $n_j=1,2,3,\ldots$ for $j=1,2$. The other two cases require antisymmetrization and symmetrization, respectively, and
are written explicitly in Appendix~A.

The matrix elements are then
\begin{equation}
{H_0}_{n,m}=\langle\psi_{n_1,n_2}|\hat{H}_{0}|\psi_{m_1,m_2}\rangle,
\end{equation}
where $n$ is shorthand for $(n_1,n_2)$, and similarly for $m$. If there were three particles then $n \equiv (n_1,n_2,n_3)$ and so on.
Here as in the one-particle case we make the matrix elements dimensionless by dividing out 
$E_1=\pi^2\hbar^2/(2m_0a^2)$, which is the single-particle ground state energy of the infinite square well; one obtains, for the distinguishable case,
\begin{equation}
{h_0}_{n,m} \equiv \frac{{H_0}_{n,m}}{E_1} =(n_1^2+n_2^2)\delta_{n_1,m_1}\delta_{n_2,m_2}.
\label{non_int_ham}
\end{equation}
Again the fermion and boson cases are given in Appendix~A.

\subsection{Contact interaction}

Without interactions the problem is of course already solved, as we are using direct products of 
the single-particle eigenstates for the basis. The
introduction of particle-particle interactions produces both diagonal and off-diagonal matrix elements,
\begin{equation}
{V_{\rm int}}_{n,m}=\langle\psi_{n_{1},n_{2}}|\hat{V}_{\rm int}|\psi_{m_{1},m_{2}}\rangle,
\label{Dirac interaction matrix elements}
\end{equation}
whose evaluation depends on the form of the interaction potential. For the contact interaction,
\begin{equation}
\hat{V}_{\rm int}=g\delta(x_{1}-x_{2}),
\label{Eq: Dirac-delta interaction potential}
\end{equation}
particles will interact with one another only if they are at the same location in space. Note that the strength of the interaction is
governed by the coefficient $g$; however, $g$ has units of energy times distance, and so a dimensionless constant $g_0$ is defined by
$g_0=g/(aE_1)$, where $a$ is the width of the well and $E_1 \equiv \hbar^2\pi^2/(2m_0 a^2)$ is the single-particle ground state energy in the absence of interactions.

Evaluating the matrix elements for this interaction yields, for distinguishable (D) particles,
\begin{eqnarray}
{V_{\rm int D}}_{n,m} &=&g\langle\phi_{n_1}(x_1)\phi_{m_1}(x_1)\phi_{n_2}(x_1)\phi_{m_2}(x_1)\rangle\nonumber \\
&=& V(n_1,n_2;m_1,m_2),
\label{con_disting}
\end{eqnarray}
where a dimensionless form of the matrix $V(n_1,n_2;m_1,m_2)$ is given by
\begin{eqnarray}
v_{n,m}&=&\frac{V(n_1,n_2;m_1,m_2)}{E_1}\nonumber\\
&=&\frac{g_0}{2}\sum_{\sigma_1,\sigma_2,s=\pm 1}\sigma_1\sigma_2\delta_{(n1-\sigma_1 n_2),s(m_1-\sigma_2 m_2)},
\label{dim_pot_mat_ele}
\end{eqnarray}
where $\sigma_1$, $\sigma_2$, and $s$ all take on values $\pm 1$, and so the complete matrix elements are then
\begin{equation}
h_{n,m}={h_0}_{n,m}+\alpha v_{n,m},
\label{Eq: matrix elements for dirac interaction ISW basis}
\end{equation}
with $\alpha = 1$. With other statistics, $\alpha = 0,1,2$, or $\sqrt{2}$, depending on the statistics, and the applicable formulas
are provided in Appendix A.

\subsection{Results}

\begin{figure}[h]
\includegraphics[width=1\columnwidth]{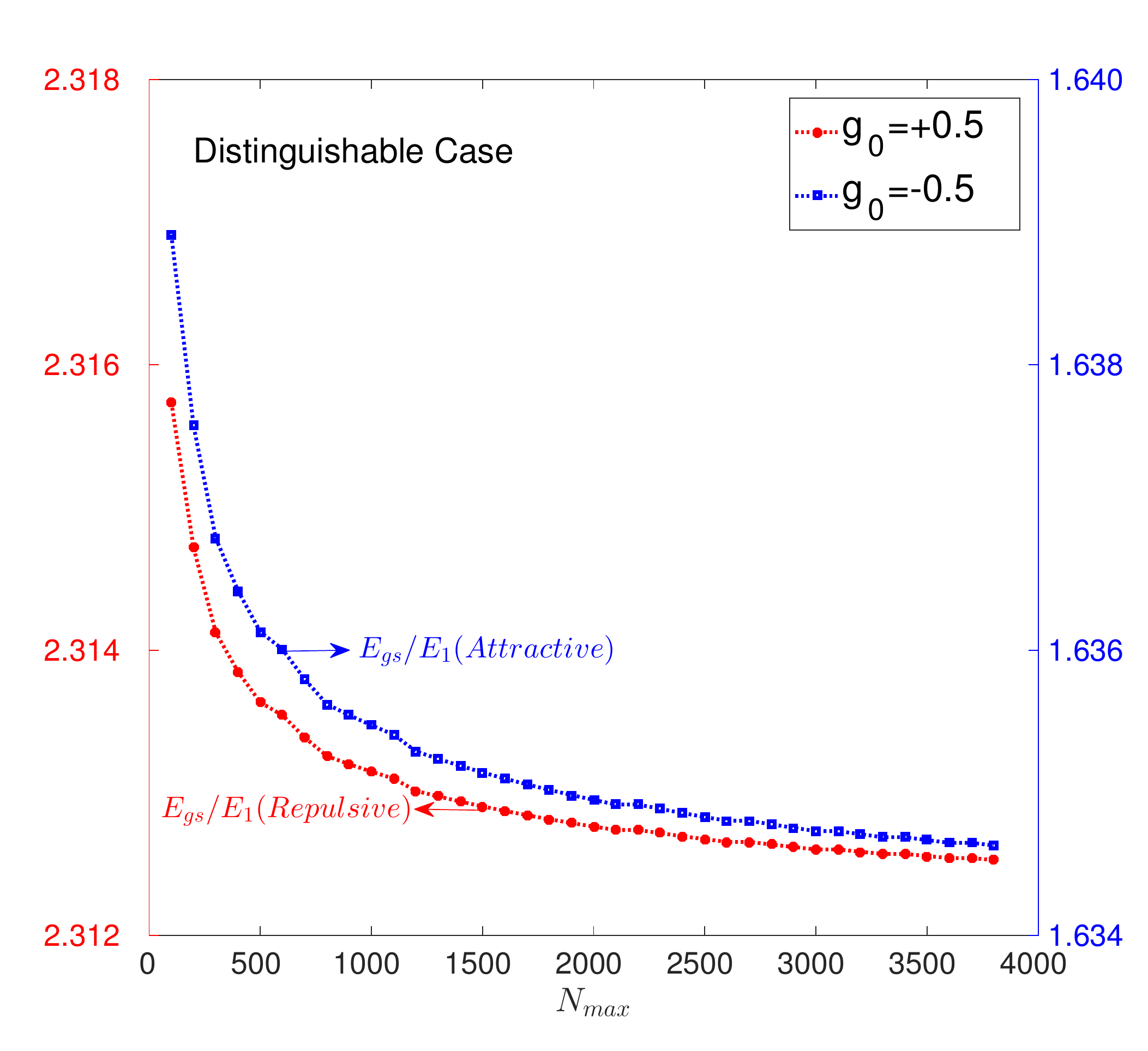}
\caption{Convergence of the ground state energy for two distinguishable particles in an infinite square well confining potential,
interacting with one another via a repulsive ($g_0 = +0.5$, lower curve, left ordinate) and an attractive ($g_0 = -0.5$, 
upper curve, right ordinate) $\delta$-function interaction. As is apparent from
the figure, results in both cases are converged at the $0.1$\% level by the time $N_{\rm max} = 4000$. For comparison the
non-interacting system has a ground state energy of $E_{\rm gs}/E_1 = 2$.}
\label{Fig: IFS only GS convergence}
\end{figure}

\begin{figure}[t]
\includegraphics[width=1\columnwidth]{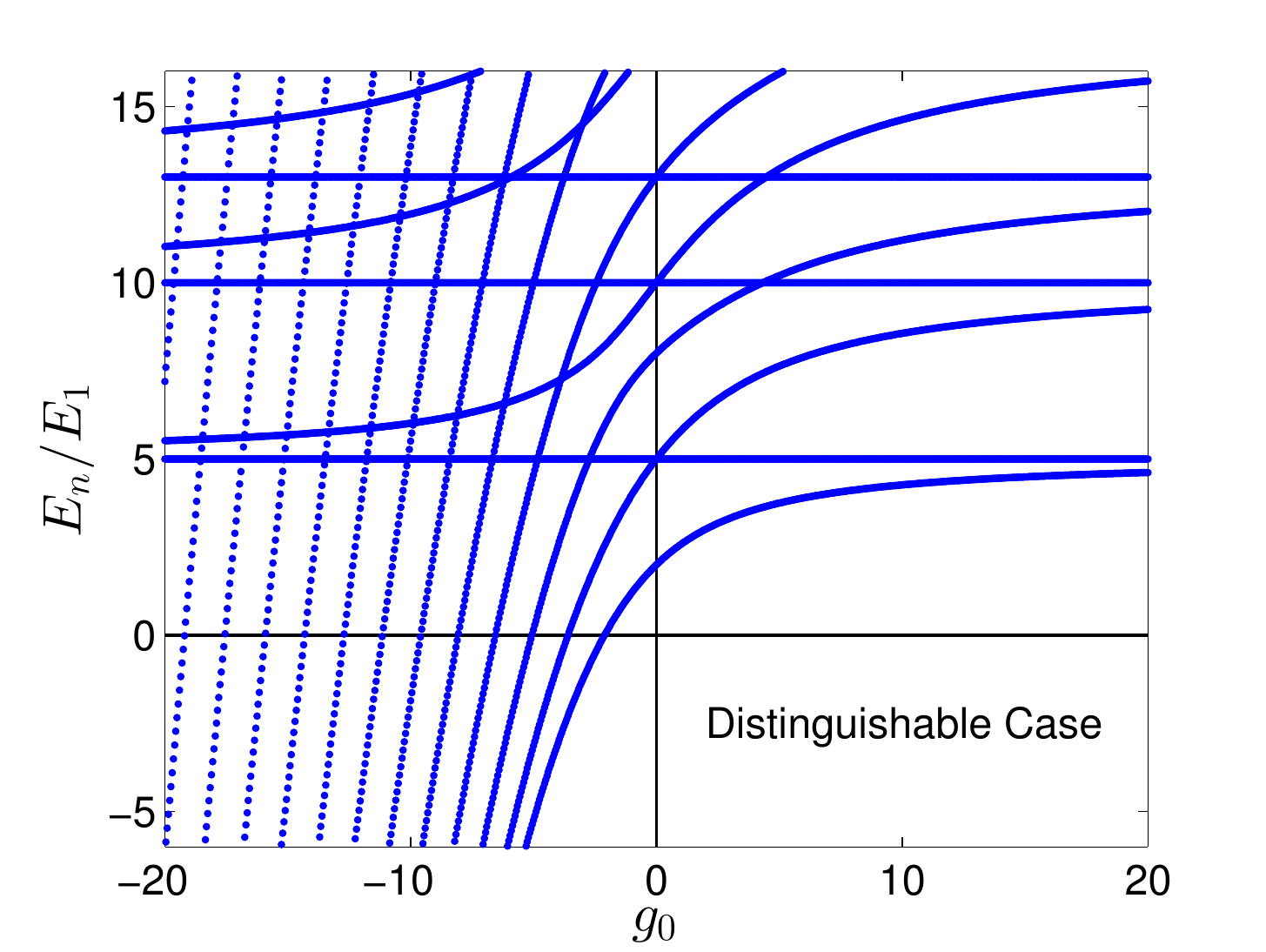}
\caption{The energy spectrum for two particles in an infinite square well, in units of $E_1 \equiv \pi^2 \hbar^2/(2m_0 a^2)$, vs
$g_0$. The two particles interact with one another through a contact potential with dimensionless strength $g_0$. These results
are obtained with $N_{\rm max} = 3856$, and are therefore completely converged on the scale of this figure. As discussed in the text,
with nonzero $g_0$ the states immediately split into fermionic (horizontal lines) and bosonic states. The fermionic states are unaffected by the interaction potential.
Note that the $g_0 = 0$ energies are the familiar $E/E_1 = 2$ for the ground state, $E/E_1 = 5$ with a degeneracy 
of $2$ for the first excited state, and so on, according to the non-interacting diagonal elements listed in Eq.~(\ref{non_int_ham}).}
\label{Fig: IFS only AGS}
\end{figure}

Using a matrix diagonalization routine,\cite{online_supplement} one can readily obtain the eigenvalues and eigenvectors of the matrix
given in Eq.~(\ref{Eq: matrix elements for dirac interaction ISW basis}), where the matrix size is 
$N_{\rm max} \times N_{\rm max}$. The states that are included for a given $N_{\rm max}$ are those whose non-interacting
(diagonal) matrix elements are below a certain prescribed value. For example, for the infinite square well basis we would
order the states according to $n_1^2 + n_2^2$, whereas for the harmonic oscillator basis (to be discussed further below) we would order them according to
$n_1 + n_2$. Figure~\ref{Fig: IFS only GS convergence} illustrates the convergence for both an attractive and a
repulsive contact potential. Not so surprisingly, convergence is not complete (to four significant figures) even for
$N_{\rm max} = 4000$, as this corresponds (roughly) to $N \approx \sqrt{4000} \approx 63$ at the single-particle level, which
was shown in Ref.~\onlinecite{marsiglio09} to be insufficient for a $\delta$-function potential for a single particle.
Nonetheless, a good qualitative picture can still be provided, as we now demonstrate for the two-particle wave function.

Figure~\ref{Fig: IFS only AGS} shows the energy spectrum obtained as a function of $g_0$. For strong attractive interactions
(negative $g_0$) there are many bound states. A bound state is defined as an eigenstate whose energy is less than zero,
since zero is the theoretical minimum energy allowed for two particles that do not interact with one another.
Note that because we are currently dealing with only two particles, the presence of an interaction is sufficient to split the
distinguishable states into two kinds: those that are symmetric and those that are antisymmetric under the operation of exchanging
the two particles. For example, if we denote the two-particle state of Eq.~(\ref{Eq: two particle basis}) by 
$|n_1 n_2\rangle$, then exchange of the two particles produces the state  $|n_2 n_1\rangle$. In the first of these particle 1 (2) is in state
$n_1$ ($n_2$), while in the second state particle 1 (2) is in state $n_2$ ($n_1$). These two states can be rearranged into symmetric,
$(|n_1 n_2 \rangle + |n_2 n_1 \rangle)/\sqrt{2}$, and antisymmetric, $(|n_1 n_2 \rangle - |n_2 n_1 \rangle)/\sqrt{2}$, combinations, and
these are the combinations that naturally emerge in the presence of an interaction. Thus, for two distinguishable particles, the eigenstates
turn out to be either fermionic (antisymmetric) or bosonic (symmetric). Even for more than two particles, there is no need
to separately calculate the energy
spectra for fermions and bosons---these emerge naturally from the spectrum for the distinguishable particle case.  Of course if one
separates these two categories at the beginning, then the Hilbert space for each category is significantly reduced compared to the
size for the indistinguishable states, and the eigenvalues and eigenvectors can be obtained more efficiently.

Returning to Fig.~\ref{Fig: IFS only AGS}, the fermionic states are readily identified by the fact that their
energies do not depend on the strength of the interaction. A contact interaction does not affect fermionic states, because fermions 
cannot be at the same place in space at the same time. Note
also that for a sufficiently large $g_0$, the two-particle wave function will develop a node when the two coordinates are equal (not shown),
so that further repulsion is immaterial. Hence, for $g_0 = 20$ for example, the ground state energy is barely increasing anymore (as a function of $g_0$). Furthermore, this saturation energy coincides with the energy of the first excited state (which is fermionic).
Similarly, for large negative values of $g_0$, various branches of the boson
energies approach (from above) the fermionic energies.  We will defer an explanation of this feature until later, after we discuss center-of-mass
excitations for particles in a harmonic oscillator potential. Then we will
illustrate that the probability associated with the wave function begins to resemble that of two fermions, so this process is
sometimes referred to as ``fermionization.''

\section{Interactions in a harmonic trap\label{sec_three}}

\subsection{Infinite-square-well basis}

We now consider the more experimentally relevant case of a harmonic oscillator confining potential, still with just two trapped, interacting particles.  As a first approach to this problem, we build on the results of the previous section and continue to use infinite-square-well basis states.  We therefore write the harmonic oscillator confining potential as
\begin{equation}
\hat{V}_{\rm conf}(x_i) = {1 \over 2} m_0 \omega^2 \left(x_i - {a \over 2}\right)^2,
\label{harm_osc}
\end{equation}
centered at the middle of the infinite square well of width~$a$ whose eigenstates will serve as our basis.  The well width $a$ must be sufficiently large that it does not affect the low-lying stationary states whose energies we wish to calculate (see Ref.~\onlinecite{marsiglio09}). 

Our basis states are again the product states of 
Eq.~(\ref{Eq: two particle basis}).  Even before 
introducing the contact interaction, these lead to the diagonal matrix elements of Eq.~(\ref{non_int_ham}) \textit{plus} additional
terms due to the confining harmonic oscillator potential.  The
non-interacting Hamiltonian can be written as $\hat{H}_0 = \hat{H}_{01} + \hat{H}_{02}$, with
\begin{equation}
\hat{H}_{0i} = - {\hbar^2 \over 2m_0}{d^2 \over dx_i^2} + {1 \over 2} m_0 \omega^2  \left(x_i - {a \over 2}\right)^2.
\label{onebodypot}
\end{equation}

\begin{figure}[t]
\includegraphics[width=1\columnwidth]{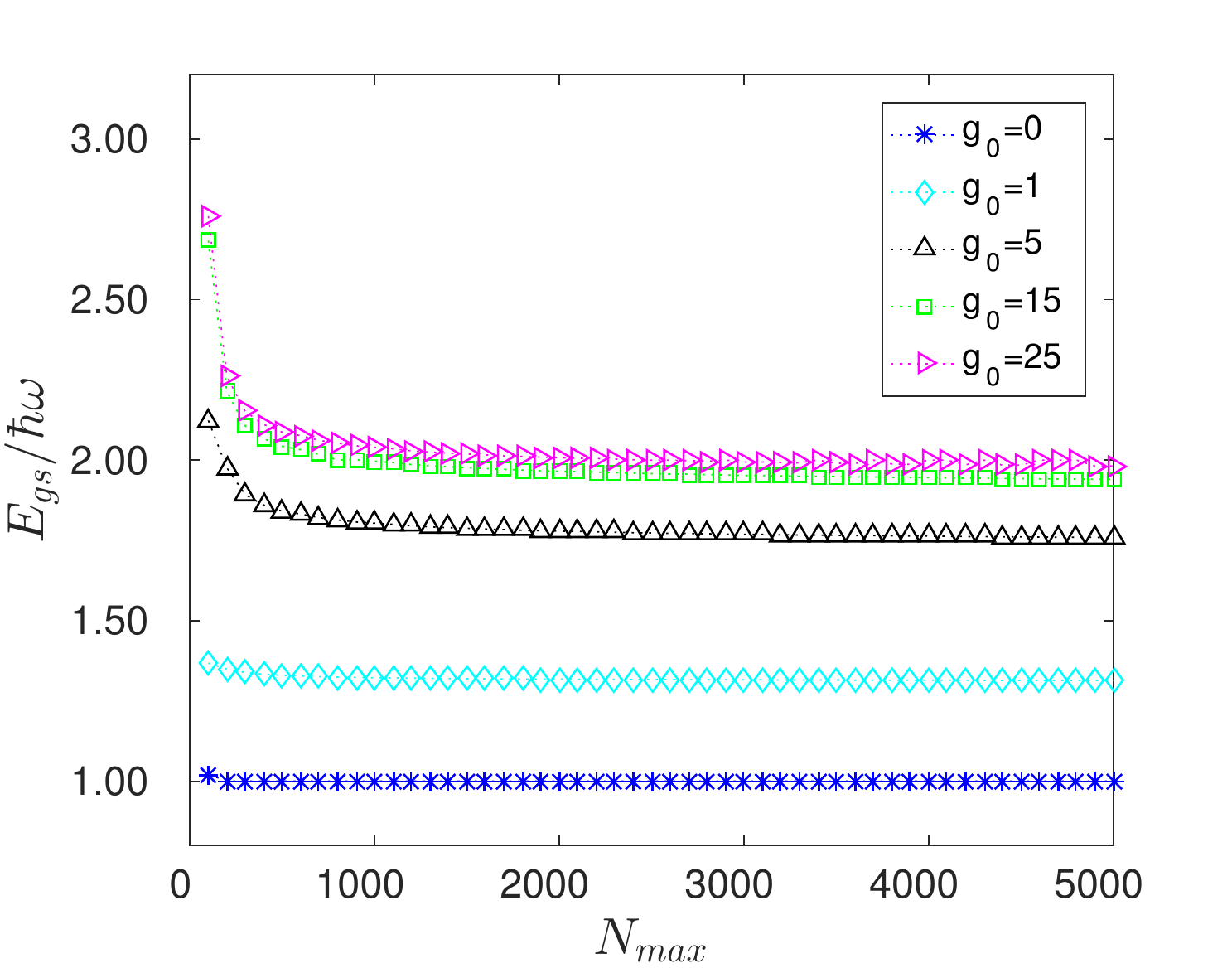}
\caption{Ground state energy convergence as a function of the number of states for two interacting particles (as discussed in the
text, in the ground state these two particles behave as bosons) in a harmonic oscillator potential embedded in an infinite square well, 
for a number of values of the dimensionless
particle-particle coupling strength, $g_0$. As
$g_0$ increases, larger and larger basis sizes are needed for convergence. Note that as $g_0$ increases the effect on energy
saturates, as `fermionization' takes place. This occurs when the repulsive interaction is strong enough to keep the two particles
apart from one another, i.e. as if they were fermions. This is clear in the relatively small change that occurs between the ground
state energy for $g_0=15$ and $g_0=25$. We have used $\rho \equiv \hbar \omega/E_1 = 50$. Note that we have normalized
the energies to $\hbar \omega$.}
\label{Fig: IFS GS convergence int}
\end{figure}

The matrix elements can then be written as
\begin{equation}
{h_0}_{n,m} \equiv \frac{{H_0}_{n,m}}{E_1} = \delta_{n_1,m_1}k_{n_2,m_2} + \delta_{n_2,m_2}k_{n_1,m_1},
\label{noninteracting}
\end{equation}
where again, following the notation of Eq.~(\ref{non_int_ham}), $n$ is shorthand for $(n_1,n_2)$, etc., and on the right-hand side, 
$k_{n_1,m_1}$ and $k_{n_2,m_2}$ are single-particle matrix elements for a particle in a harmonic
oscillator potential.  These matrix elements have the form\cite{marsiglio09}
\begin{eqnarray}
k_{n_1,m_1}&=&\delta_{n_1,m_1}\bigg[n_1^2+\frac{\pi}{48}\bigg(\frac{\hbar\omega}{E_1}\bigg)^2\biggl(1-\frac{6}{(n_1\pi)^2}\biggr)\bigg]
\nonumber \\
&&\qquad+(1-\delta_{n_1,m_1})\bigg(\frac{\hbar\omega}{E_1}\bigg)^2\eta_{n_1,m_1},
\label{Eq. matrix elements of HO in ISW}
\end{eqnarray}
where
\begin{equation}
\eta_{n_1,m_1}=\frac{(-1)^{n_1+m_1}+1}{4}\bigg(\frac{1}{(n_1-m_1)^2}-\frac{1}{(n_1+m_1)^2}\bigg),
\label{Eq: ISW/HO g}
\end{equation}
and similarly for $k_{n_2,m_2}$. The only remaining piece is the matrix element corresponding to the contact interaction,
and it is the same as in Eq.~(\ref{dim_pot_mat_ele}).  Combining this equation with
Eq.~(\ref{noninteracting}), the matrix elements are given by
\begin{equation}
h_{n,m} = h_{0_{n,m}} + v_{n,m}.
\label{matrix_elements}
\end{equation}

\subsection{Results}

So how well does this work? 

\begin{figure}[b]
\includegraphics[width=1\columnwidth]{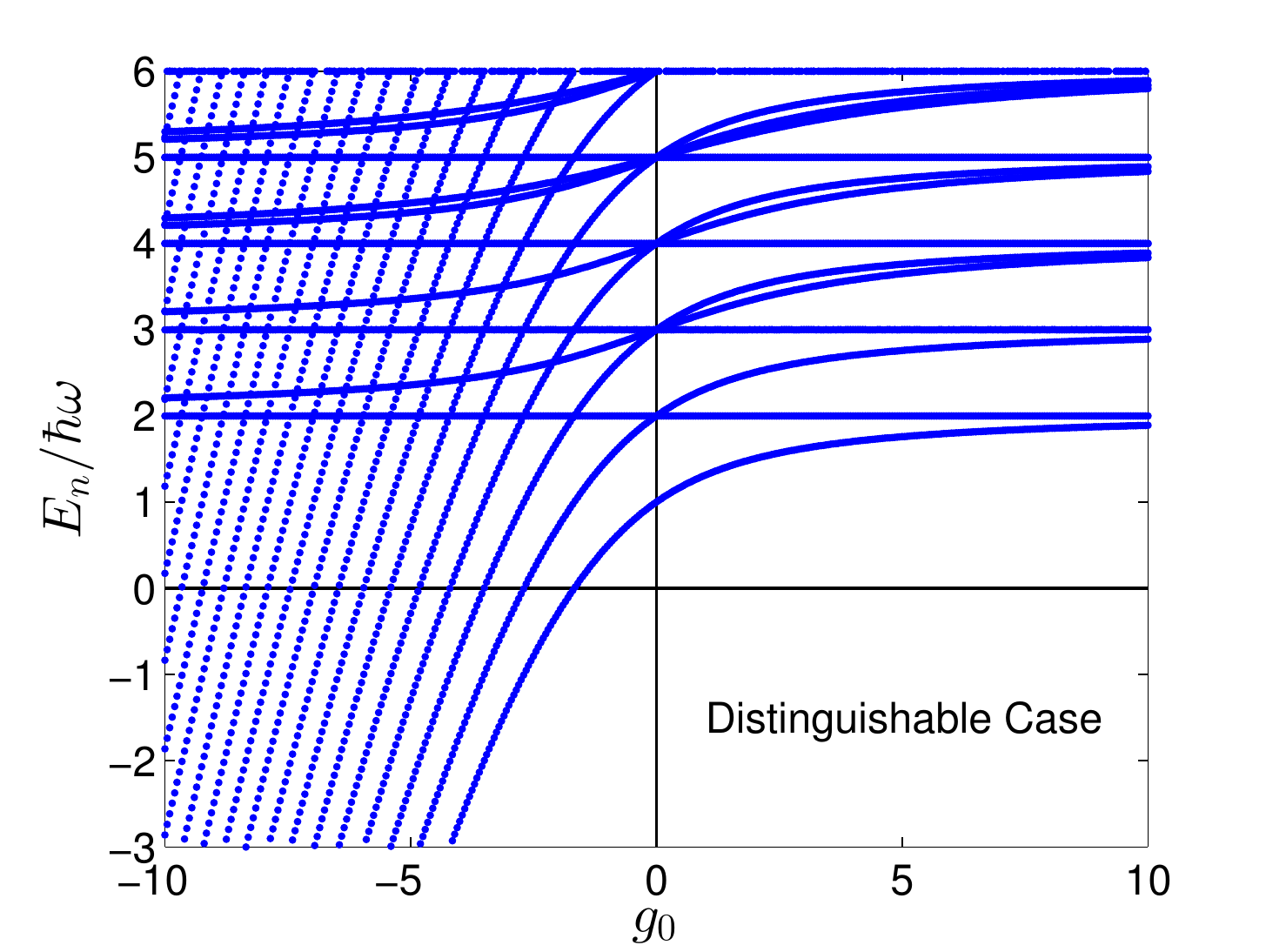}
\caption{Energy levels (normalized to $\hbar \omega$) as a function of the strength of the particle-particle contact interaction, $g_0$ for two particles with
mass $m_0$ each in a harmonic trap with frequency $\omega$. As was the case with 
the infinite square well trap, the fermionic states are easily identifiable as the horizontal lines that are unaffected by the interaction. Also,
as the interaction strength increases, the boson state energies approach the energy of the fermion state above it (`fermionization') as was
the case with the infinite square well trap. For this figure we used $\rho \equiv \hbar \omega/E_1 = 50$ and $N_{\rm max} = 5029$.}
\label{Fig: IFS HO AGS}
\end{figure}

Figure~\ref{Fig: IFS GS convergence int} shows the ground-state energy as a function of the number
of basis states used; the impact of the contact interaction clearly slows down
the convergence as a function of the number of basis states, especially when $g_0$ is large. Convergence will also
depend on the width of the square well used; in this and subsequent figures we have used a width such that the dimensionless parameter
$\rho \equiv \hbar \omega/E_1 = 50$.  This value represents a sufficiently wide well that the walls of the well do not affect the
results for the ground state (and for many excited states as well).
Clearly there is a difference
between the bosonic vs fermionic (not shown) eigenstates, since the interaction is effectively absent in the latter case. Nonetheless,
we do achieve convergence to a given accuracy, and these results will serve as a benchmark for more refined calculations below. 
Note that for large $g_0$ the energy barely increases as $g_0$ is increased further, for reasons discussed at the end of Sec.~\ref{sec_two}.

Figure~\ref{Fig: IFS HO AGS} shows the energy levels as a function of $g_0$.  The behavior is qualitatively similar to that of the infinite square well, shown in Fig.~\ref{Fig: IFS only AGS}. There are sets of states whose energies do not change
as a function of $g_0$---these are the fermionic states for which the contact interaction remains invisible, since two fermions
cannot occupy the same point in space.  Other (bosonic) states are affected by this interaction; in particular for negative values of
$g_0$ there is an increasing number of bound states as $g_0$ decreases. The origin of these will be clarified below. Furthermore,
as the interaction strength increases, the energy of each bosonic state approaches the energy of the fermion state above it --- this is
the phenomenon of ``fermionization'' referred to above, and the bosons, due to the large repulsion between them,  behave somewhat like fermions.

In what follows we will take two additional steps to redo the calculation just presented. 
First we will adopt a basis set that consists of products of the eigenstates of the single-particle
harmonic oscillator problem. These basis states are far more natural for the harmonic oscillator confining potential. Second, we 
will utilize so-called center-of-mass variables to simplify the problem from an 
$N$-body to an $(N-1)$-body problem.  This will have a more significant impact when $N$ is small, and we will explicitly examine $N=2$ for illustration purposes.

\subsection{Harmonic oscillator basis}  

Unlike with the infinite square well basis, if we are to use the products of the single particle harmonic oscillator eigenstates as
basis states, then there is no point to centering the harmonic oscillator confining potential away from $x=0$. 
Then the single particle problem is solved by the usual wave functions,
\begin{equation}
\phi_n(x)=\bigg(\frac{m_0 \omega}{\pi\hbar}\bigg)^{1/4}\frac{1}{\sqrt{2^nn!}}\,H_n\bigg(\sqrt{\frac{m_0 \omega}{\hbar}}x\bigg)\exp\Bigl({-\frac{m_0 \omega}{2\hbar}x^2}\Bigr),
\label{Eq: one particle HO basis}
\end{equation}
with eigenenergies
\begin{equation}
\epsilon_n=\textstyle\hbar\omega(n+\frac{1}{2}),
\label{Eq: E for single particle in HO well}
\end{equation}
where $n= 0,1,2,\ldots$ is a whole number and $H_n(z)$ is the usual Hermite polynomial.\cite{remark2}

While this basis is more complicated, it has the advantages that (i) no ``embedding'' potential like the infinite
square well is needed, and (ii) no effort is required for the non-interacting case.  Let us define dimensionless matrix elements this time by dividing all energies by $\hbar \omega$, i.e., ${h_0}_{n,m} \equiv {H_0}_{n,m}/\hbar \omega$.  Then, for the distinguishable case,
we have simply
\begin{equation}
{h_0}_{n,m}=(n_1+n_2+1)\,\delta_{n_1,m_1}\delta_{n_2,m_2}.
\label{Eq:Eigenenergies for dist/indist particles}
\end{equation}
Only the diagonal elements of this matrix are nonzero, because the basis functions are the 
exact solution to the two-particle system without interactions.

With the interaction $V(x_1 - x_2) = g \delta(x_1 - x_2)$,
the required matrix elements, using the basis of product states of
the single particle states in Eq.~(\ref{Eq: one particle HO basis}), are
\begin{eqnarray}
{v_{\rm int}}_{n,m} & \equiv & {{V_{\rm int}}_{n,m} \over \hbar \omega} \nonumber \\
&=&g\int_{-\infty}^{\infty}\phi^\ast_{n_1}(x_1)\phi^\ast_{n_2}(x_1)\phi_{m_1}(x_1)\phi_{m_2}(x_1)\,dx_1.\phantom{aaaa}
\label{vint}
\end{eqnarray}
Using a dimensionless coupling constant $g_{\rm ho}=\sqrt{{m_0 \omega/\hbar}}\,{g/(\hbar\omega)}$,
we require
\begin{equation}
{v_{\rm int}}_{n,m} = g_{\rm ho}c\int_{-\infty}^{\infty}H_{n_1}(z)H_{n_2}(z)H_{m_1}(z)H_{m_2}(z)e^{-2z^2}dz,
\label{Eq: integral Dirac-delta matrix element}
\end{equation}
where 
\begin{equation}
c\equiv \frac{1}{\pi}2^{-(n_1+n_2+m_1+m_2)/2}(n_1 !\,n_2 !\,m_1 !\,m_2 !)^{-1/2}
\label{c_defn}
\end{equation}
is a constant with respect to the integration variable $z \equiv x_1\sqrt{m_0 \omega/\hbar}$. Note that for a given interaction strength $g$ that is independent of the confining potential, the two dimensionless coupling strengths are
related through $g_{\rm ho}/g_0 = \pi \sqrt{E_1/(2 \hbar \omega)} = \pi /\sqrt{2\rho}$.

The integral in Eq.~(\ref{Eq: integral Dirac-delta matrix element}) can be done numerically. However, the integrand will be highly
oscillatory as the quantum numbers increase, and it is worthwhile to examine alternative procedures.\cite{deuretzbacher08} First, because
the Hermite polynomials have a definite parity, we have
\begin{equation}
v_{{\rm int}_{n,m}}=\begin{cases}
		g_{\rm ho}c\int_{-\infty}^{\infty}H_{n_2}(z)H_{m_2}(z)H_{n_1}(z)H_{m_1}(z)e^{-2z^2}dz &\\
		\quad \quad \quad \text{if}\quad (n_1+n_2+m_1+m_2)\quad \text{is even} \\
		0  \ \ \quad \quad \text{if}\quad (n_1+n_2+m_1+m_2)\quad \text{is odd},\\
		\end{cases}\\	
\label{Eq:Dirac parity thing}
\end{equation}
so only half the integrals are required.

\subsubsection{The ``brute force'' solution}

The nonzero integrals can be done analytically by 
using the expansion
\begin{equation}
{\displaystyle H_{n}(z)=\sum_{s=0}^{[n/2]}(-1)^{s}(2z)^{(n-2s)}\frac{n!}{(n-2s)!s!}},
\label{Eq: Hermite sum}
\end{equation} 
where $[n/2] = n/2$ if $n$ is even and $[n/2] = (n-1)/2$ if $n$ is odd. Using this expression
in Eq.~(\ref{Eq: integral Dirac-delta matrix element}) and performing the integral leaves us with
\begin{equation}
{v_{\rm int}}_{n,m}=g_{\rm ho}c\sqrt{\frac{\pi}{2}}\sum_{s} {1 \over 2^{n'}}f_{n_1,s_1}f_{n_2,s_2}f_{m_1,\sigma_1}f_{m_2,\sigma_2}
\frac{(2n')!}{n'!},
 \label{Eq: The brute force method (Dirac)}
\end{equation}
where
\begin{equation}
f_{n_1,s_1}=(-1)^{s_1}\frac{n_1!}{(n_1-2s_1)!s_1!}
\label{Eq: brute force f}
\end{equation}
and 
\begin{equation}
n'=\frac{1}{2}(n_1+n_2+m_1+m_2)-(s_{1}+s_{2}+\sigma_{1}+\sigma_{2})
\label{nprime}
\end{equation}
and
\begin{equation}
\sum_{s} \equiv \sum_{s_{1}=0}^{[\frac{n_{1}}{2}]}\,\sum_{s_{2}=0}^{[\frac{n_{2}}{2}]}\,\sum_{\sigma_{1}=0}^{[\frac{m_{1}}{2}]}\,\sum_{\sigma_{2}=0}^{[\frac{m_{2}}{2}]}.
\label{sums}
\end{equation}
As mentioned before, this method is computationally taxing, and despite all the integrals set to zero due to parity, this embedded quadruple sum is the reason why it will still take considerable time to compute these matrix elements, either numerically, or with the expansion
given in Eq.~(\ref{Eq: Hermite sum}).

\subsubsection{The Wang solution}

An alternative solution uses an identity due to Wang.\cite{wang09} Details of the derivation are given in Appendix~B.  Here we outline the key ideas. 

To evaluate the integral in Eq.~(\ref{Eq: integral Dirac-delta matrix element}), we exploit the orthogonality of the Hermite polynomials, expressed by
\begin{equation}
\int_{-\infty}^{\infty}e^{-z^2}H_\ell(z)H_{\ell'}(z)\,dz=\delta_{\ell\ell'} 2^\ell \ell!\sqrt{\pi}.
\label{Eq: HO basis orthogonality}
\end{equation}
Since $2z^2$ appears in the exponential in Eq.~(\ref{Eq: integral Dirac-delta matrix element}), 
and since Hermite polynomials are just polynomials,
one can write the product of two Hermite polynomials with argument $z$ as a linear combination of single Hermite polynomials
(necessarily of higher order) with argument $\alpha z$, where $\alpha$ is any constant. In this case, because of the form of the exponential, we 
choose $\alpha = \sqrt{2}$.  That is, we write
\begin{equation}
H_j(z)H_k(z)=\sum_{r=0}^{j+k}a_r(j,k)H_r(\sqrt{2}z).
\label{Eq: Wang's trick}
\end{equation}
Then the general integral we require can be written
\begin{equation}
\begin{split}
I=& \ I(j,k,p,q)\\
=&\int_{-\infty}^{\infty}H_{j}(z)H_{k}(z)H_{p}(z)H_{q}(z)e^{-2z^{2}}dz\\
=&\sum_{\ell=0}^{j+k}\sum_{\ell'=0}^{p+q}a_\ell(j,k) a_{\ell'}(p,q)\\
&\qquad \int_{-\infty}^{\infty}H_\ell (\sqrt{2}z)H_{\ell'}(\sqrt{2}z) e^{-(\sqrt{2}z)^2}dz\\
=&\sum_{\ell=0}^{j+k}\sum_{\ell'=0}^{p+q}a_{\ell}(j,k) a_{\ell'}(p,q)\sqrt{\frac{\pi}{2}}2^\ell \ell!\delta_{\ell,\ell'}\\
\text{} =&\sum_{\ell=0}^{\ell_{\rm max}}a_\ell(j,k) a_\ell(p,q)\sqrt{\frac{\pi}{2}}2^\ell \ell!
\end{split}
\label{Eq:hermite simplify}
\end{equation}
where $\ell_{\rm max} = \textrm{min}\{j+k,p+q\}$, and $\ell$ in the last line is over even (odd) numbers only if $j+k$ is even (odd). Note that
if $j+k$ is even (odd) then $p+q$ is also even (odd) for all nonzero integrals. Note also that there is a freedom of choice for the 
$i$, $j$, $p$, and $q$ at the start of Eq.~(\ref{Eq:hermite simplify}). In particular it makes sense to pair the two lowest values of these
four indices (quantum numbers) so that the upper limit $\ell_{\rm max}$ at the end of Eq.~(\ref{Eq:hermite simplify})
will be the smallest possible number. It remains to determine $a_\ell(j,k)$; the details are in the appendix. The result is
\begin{eqnarray}
a_\ell(j,k)=&&\frac{j!k!}{2^{(j+k)/2}}\frac{(-1)^{(j+k-\ell)/2}}{(\frac{j+k-\ell}{2})!\ell!}\times \nonumber \\
&&\sum_{u=\textrm{max}(\ell-j,0)}^{\textrm{min}(k,\ell)} (-1)^{k-u} C_{k-u}^{j+k-\ell}C_u^\ell,
\end{eqnarray}
where 
\begin{equation}
C_u^\ell \equiv {\ell ! \over (\ell - u)!\, u !}.
\label{comb}
\end{equation}
In summary, following Wang\cite{wang09} allows us to replace the four embedded sums with a single sum requiring two individual sums 
(in $a_\ell(j,k)$).  For large matrices, on a personal computer, this reduces a calculation that would have taken many days to a few minutes.

\subsubsection{Results}

\begin{figure}[t]
\includegraphics[width=1\columnwidth]{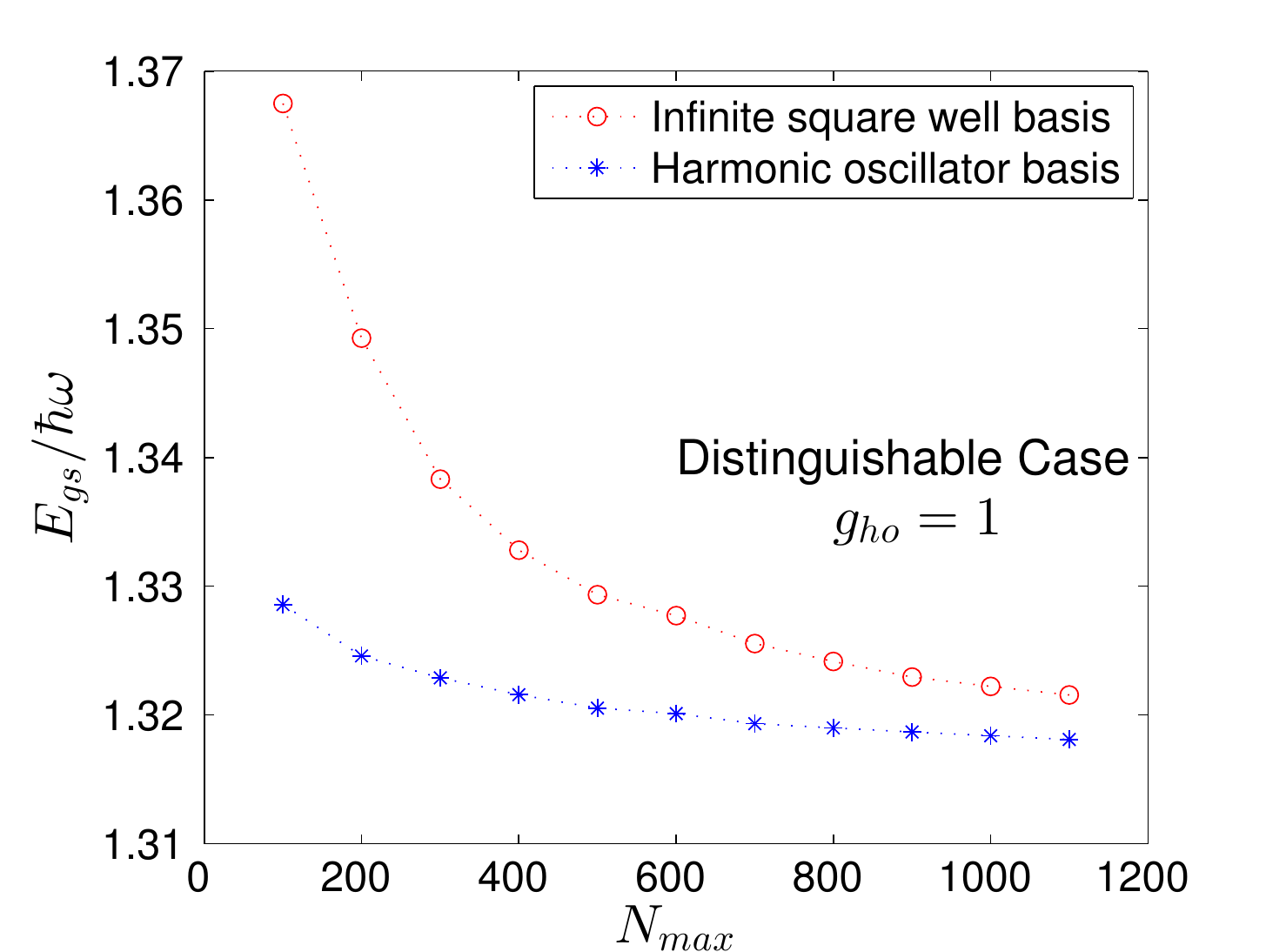}
\caption{Comparison of the rate of convergence between the infinite square well and harmonic oscillator basis sets. As expected, the harmonic oscillator basis converges more quickly as a function of basis size than the infinite square well basis. Note that with $g_{\rm ho} = 1$ then
$g_0$ has to be determined for a given infinite square well width (i.e. given $\rho \equiv \hbar \omega /E_1$) through the relation
$g_0 = g_{\rm ho}  \sqrt{2\rho/\pi}$ with $\rho = 50$. Results for the ground state do not depend on this width, when chosen to
be sufficiently large.}
\label{Fig: ISWvHO GS convergence}
\end{figure}

The results of these calculations are of course the same as those shown in Fig.~\ref{Fig: IFS HO AGS} for an infinite square well basis. 
Here, however, results will be valid for all states, whereas in the infinite square well basis the more energetic states (not shown in 
Fig.~\ref{Fig: IFS HO AGS}) can ``feel'' the walls of the square well, and hence are no longer
solutions for the harmonic trap alone. However, convergence as a function of basis state size will differ, so a comparison is provided in 
Fig.~\ref{Fig: ISWvHO GS convergence} for the ground state. As this figure shows, the harmonic oscillator basis leads to faster 
convergence as a function of basis size. 

We could move on to more particles at this stage, and the path should be clear.
First, however, we examine a simplification that
can be made to remove the center-of-mass motion for any number of particles in a harmonic trap, and illustrate the procedure for
two particles.

\subsection{Removal of the center of mass}

For $N$ particles one can use so-called Jacobi coordinates (see, for example, Ref.~\onlinecite{liu10}); 
these include the center-of-mass coordinate,
\begin{equation}
x_{\rm cm} \equiv \frac{x_1 + x_2 + \cdots + x_N}{N},
\label{r_cofm}
\end{equation}
and relative coordinates
\begin{equation}
{x_r}_i \equiv \sqrt{i - 1 \over i}\left(x_i - {1 \over i-1} \sum_{k=1}^{i-1} x_k\right),
\label{r_i}
\end{equation}
for $i\ge2$.

Use of these coordinates allows the center-of-mass variable to be removed, leaving a problem in $N-1$ variables. 

For two particles, one can utilize this same transformation, or use a slight variant with Jacobian equal to unity. This is
accomplished by using a center-of-mass coordinate $x_c$ and a relative
coordinate $x_r$, defined by
\begin{equation}
x_{c}=\frac{x_1+x_2}{2}, \qquad x_{r}=x_1-x_2.
\label{cofm_rel_two}
\end{equation}
Proceeding with this transformation, the Hamiltonian
\begin{equation}
\hat{H}=\frac{\hbar^2}{2m_0}\bigg(\frac{d^2}{dx_1^2}+\frac{d^2}{dx_2^2}\bigg)+\frac{1}{2}m_0 \omega^2(x_1^2+x_2)^2+g\delta(x_1-x_2)
\label{Eq: Not COM hamiltonian}
\end{equation}
becomes
\begin{equation}
\begin{split}
\hat{H}&=-\frac{\hbar^2}{2m_c}\frac{d^2}{dx_{c}^2}+\frac{1}{2}m_c\omega^2x_{c}^2\\
&-\frac{\hbar^2}{2\mu}\frac{d^2}{dx_{r}^2}+\frac{1}{2}\mu\omega^2x_{r}^2+g\delta(x_{r})\\
&=\hat{H}_{cm}+\hat{H}_{rel}
\end{split}
\end{equation}
where $m_c=2m_0$ and $\mu=m_0/2$ are the total mass and the reduced mass, respectively. 
The resulting Schr\"odinger equation is separable, and the solution to the center-of-mass Hamiltonian is just the 
solution to a single particle harmonic oscillator with mass $m_c$ and frequency $\omega$, with eigenvalues 
$E_{n_c}^\textrm{cm}=\hbar\omega(n_c+1/2)$ with $n_c = 0,1,2,\ldots\,$. 

\begin{figure}[t]
\includegraphics[width=1\columnwidth]{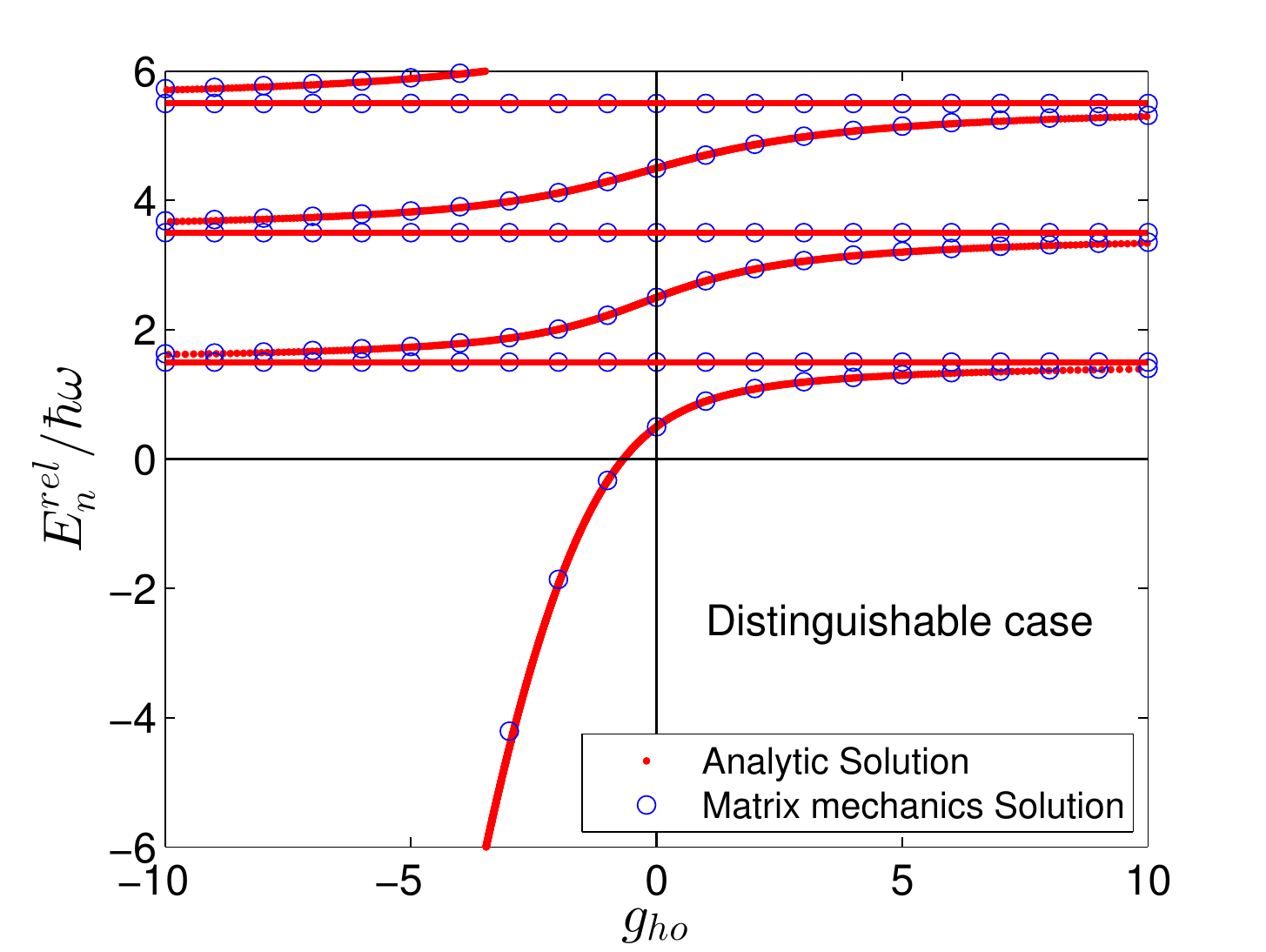}
\caption{Normalized energy levels vs. dimensionless contact potential strength, $g_{\rm ho}$, for two interacting particles in a
harmonic trap. The curves in red are analytical solutions, obtained through Eq.~(\ref{e_anal}).
The points in blue are determined by solving the matrix diagonalization problem for the effective single 
particle problem (center-of-mass motion removed), as
defined in Eq.(\ref{harm1}), in the harmonic oscillator basis. Alternatively, one could solve this problem in the infinite square well
basis (not shown); the results are identical. As is apparent from the figure, these two results are in essentially perfect agreement 
with one another. We used $N_{\rm max} = 2000$. }
\label{Fig: Erel AGS}
\end{figure}

This change of variables leaves the Hamiltonian in the relative coordinate, which then describes a one-body problem of a 
particle of mass $\mu = m_0/2$
in a potential consisting of
a harmonic oscillator potential plus a $\delta$-function potential, both centered at the origin. Viewed in this way, since the potential is
an even function of $x_r$, solutions are either even
or odd in $x_r$, i.e., they are either symmetric  or antisymmetric in $(x_1,x_2)$, respectively. Solutions that are odd do not ``see'' the
$\delta$-function potential.  Therefore these are the usual (odd) solutions for the harmonic oscillator, with energies
$E_{n}^\textrm{rel}=\hbar\omega(n_r+1/2)$, with $n_r = 0,1,2,\ldots\,$. In combination with the center-of-mass solutions, these solutions
represent the fermion solutions to the problem, since a solution that is odd in $x_r$ is antisymmetric in $x_1$ and $x_2$ 
(see Eq.~(\ref{cofm_rel_two}) and note that the center-of-mass solution is always symmetric in $x_1$ and $x_2$).

\begin{figure}[b]
\includegraphics[width=1\columnwidth]{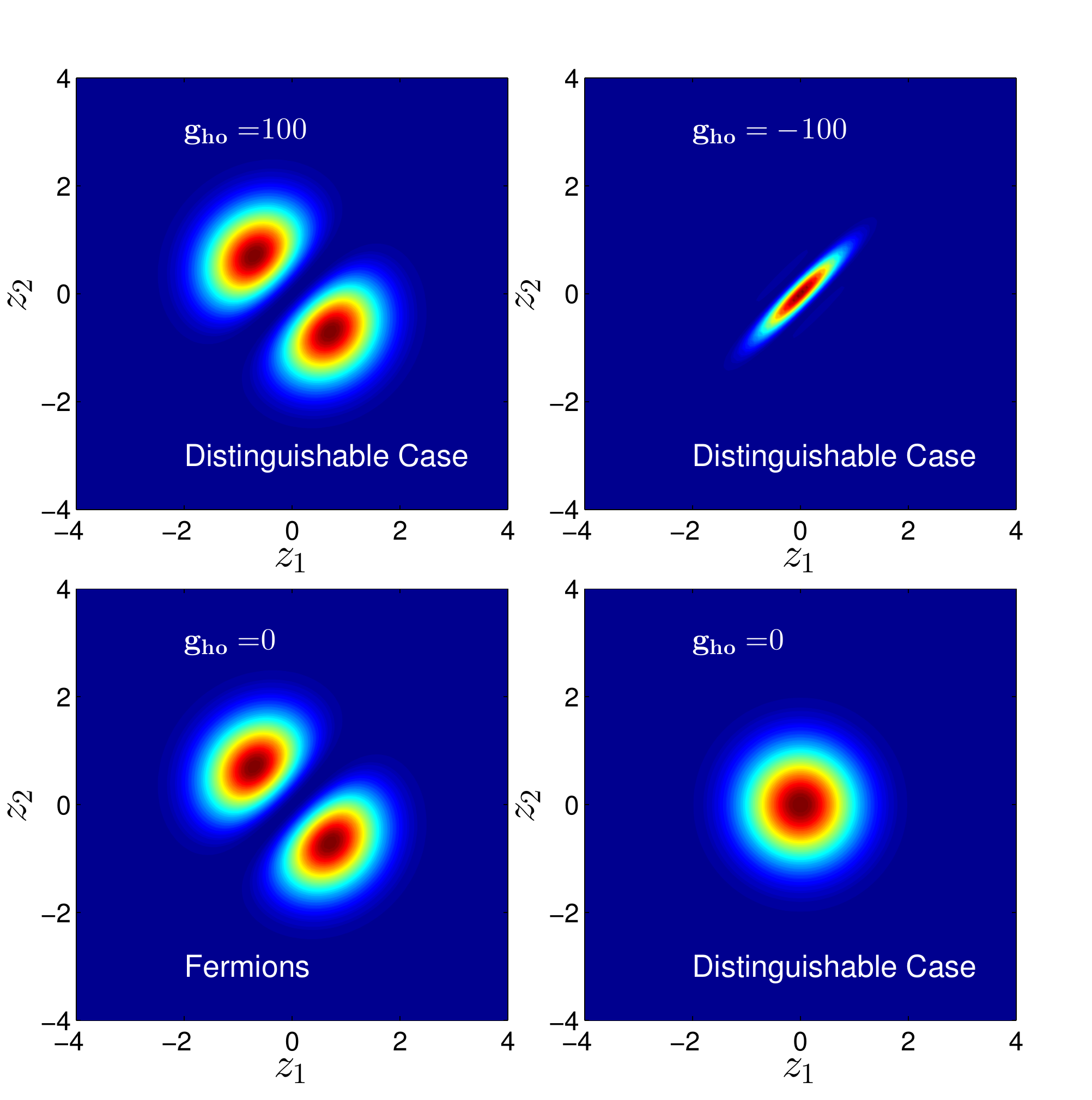}
\caption{A contour plot of the ground state two-particle wave function, $|\psi(x_1,x_2)|^2$, as a function of the dimensionless
positions $z_1$ and $z_2$ for (i) top left panel, $g_{\rm ho} = +100$, (ii) top right panel, $g_{\rm ho} = -100$, (iii) bottom
left panel, fermion case ground state, and (iv) bottom right panel, the non-interacting case. In practice these are all achieved using the distinguishable
basis states for the various particle-particle interaction strengths shown, except the fermion ``ground state'' corresponds to the
first excited state (for any coupling strength, since it is independent of coupling strength). Note the degree of overlap in the two
particles in the strongly attractive case (ii), as well as the near agreement of the strongly repulsive case (which is bosonic) (i) with the fermion case (iii), illustrating the `fermionization' of the former.}
\label{Fig: Wavefun comp}
\end{figure}

For the boson solutions, we require solutions that are even in $x_r$. These were first determined analytically less than 20 years 
ago.\cite{busch98} See also Refs.~\onlinecite{patil06,viana-gomes11}.
The resulting eigenvalues are given by the implicit equation
\begin{equation}
{g_{\rm ho}} = \left(e_n - {1\over 2}\right) {\Gamma \left({3 \over 4}-{e_n \over 2}\right) \over \Gamma \left({5 \over 4} - {e_n \over 2}\right)}
\label{e_anal}
\end{equation}
where $g_{\rm ho}$ is defined as earlier (with the particle mass, \textit{not} the reduced particle mass), 
$e_n \equiv E_{n}^\textrm{rel}/(\hbar\omega)$, and the $\Gamma$ functions
are the usual ones.\cite{abramowitz72,nist10} This expression agrees with those obtained in 
Refs.~\onlinecite{busch98,patil06,viana-gomes11} and allows for easy evaluation of $g_{\rm ho}$ in terms of $e_n$, although
the latter is usually plotted as a function of the former.

\begin{figure}[b]
\includegraphics[width=1\columnwidth]{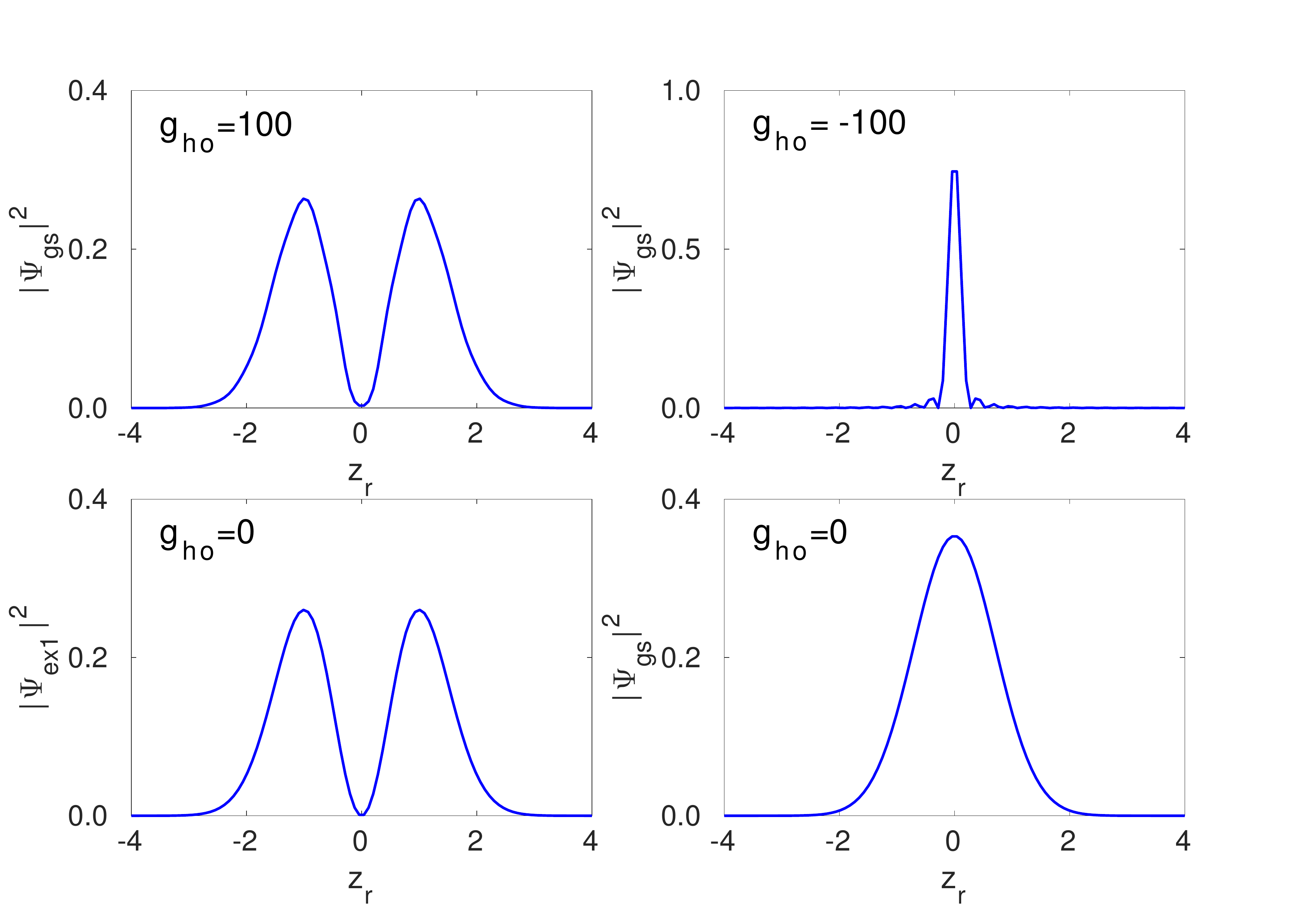}
\caption{Same 4 cases as in Fig.~\ref{Fig: Wavefun comp}, with now the relative wave function vs. the relative coordinate 
$x_r$. Again note the close resemblance of the (boson) ground state in the upper left panel with the (non-interacting) fermion
ground state in the lower left panel.}
\label{Fig: Wavefun comp_relative}
\end{figure}

Alternatively, and more straightforwardly, we solve the problem numerically, using matrix mechanics with a harmonic oscillator 
basis, i.e., the one given by 
Eq.~(\ref{Eq: one particle HO basis}).
Then the matrix elements are simply
\begin{equation}
h_{n,m} = \delta_{n,m} \left(n+{1 \over 2}\right) + {g_0 \over \sqrt{2\pi}} f_n f_m,
\label{harm1}
\end{equation} 
where 
\begin{equation}
f_n=\begin{cases}
		0 & \text{for $n$ odd}, \\
		\displaystyle {(-1)^{n/2} \over (n/2)!} \sqrt{n! \over 2^n} & \text{for $n$ even},
		\end{cases}\\
\label{f_defn}
\end{equation}
and for ease of computation one can use the recursion relation $f_{n+2} = \sqrt{(n+2)/(n+1)} f_n/2$ to compute $f_n$ for large $n$.

The relative energies are shown in Fig.~\ref{Fig: Erel AGS}. Note that for sufficiently negative $g_{\rm ho}$ there is only one
bound state. This is compatible with the picture shown in Fig.~\ref{Fig: IFS HO AGS} because
the multiple bound states illustrated there arise due to the fact that the one bound state shown here, in Fig.~\ref{Fig: Erel AGS},
can be excited through center-of-mass excitations ($n_c \ne 0$), and still remain a bound state. In fact, starting with the energies
shown in Fig.~\ref{Fig: Erel AGS}, if one adds the center-of-mass energies, $E_{n_c}^\textrm{cm} = \hbar \omega (n_c + 1/2)$, 
for $n_c = 1,2,3,\ldots\,$, then the results are in excellent agreement with those shown in Fig.~\ref{Fig: IFS HO AGS}.
These center-of-mass excitations
have varying importance, depending on the circumstance. Here, in a harmonic trap, they need to be accounted for, whereas, in
the context of a nucleus with many nucleons, they are regarded as spurious, and correspond to the motion of the entire
nucleus through space.

Note that the odd-parity solutions have energies that are independent of $g_{\rm ho}$ (horizontal lines).
As emphasized earlier, for large values of $|g_{\rm ho}|$, the boson energies tend to the fermion energies, the process already referred to
as ``fermionization.''  For large positive values of $g_{\rm ho}$ the physical interpretation is clear: a very strong repulsion between
particles mimics the Pauli exclusion principle, and the bosons behave as fermions. For large negative values of $g_{\rm ho}$ the boson ground state is a strongly peaked $\delta$-function-like wave function. All the excited states are orthonormal to the ground state,
and will have structure that approaches a node at the origin to achieve this, as we illustrate in the remainder of this section.

\begin{figure}[b]
\includegraphics[width=1\columnwidth]{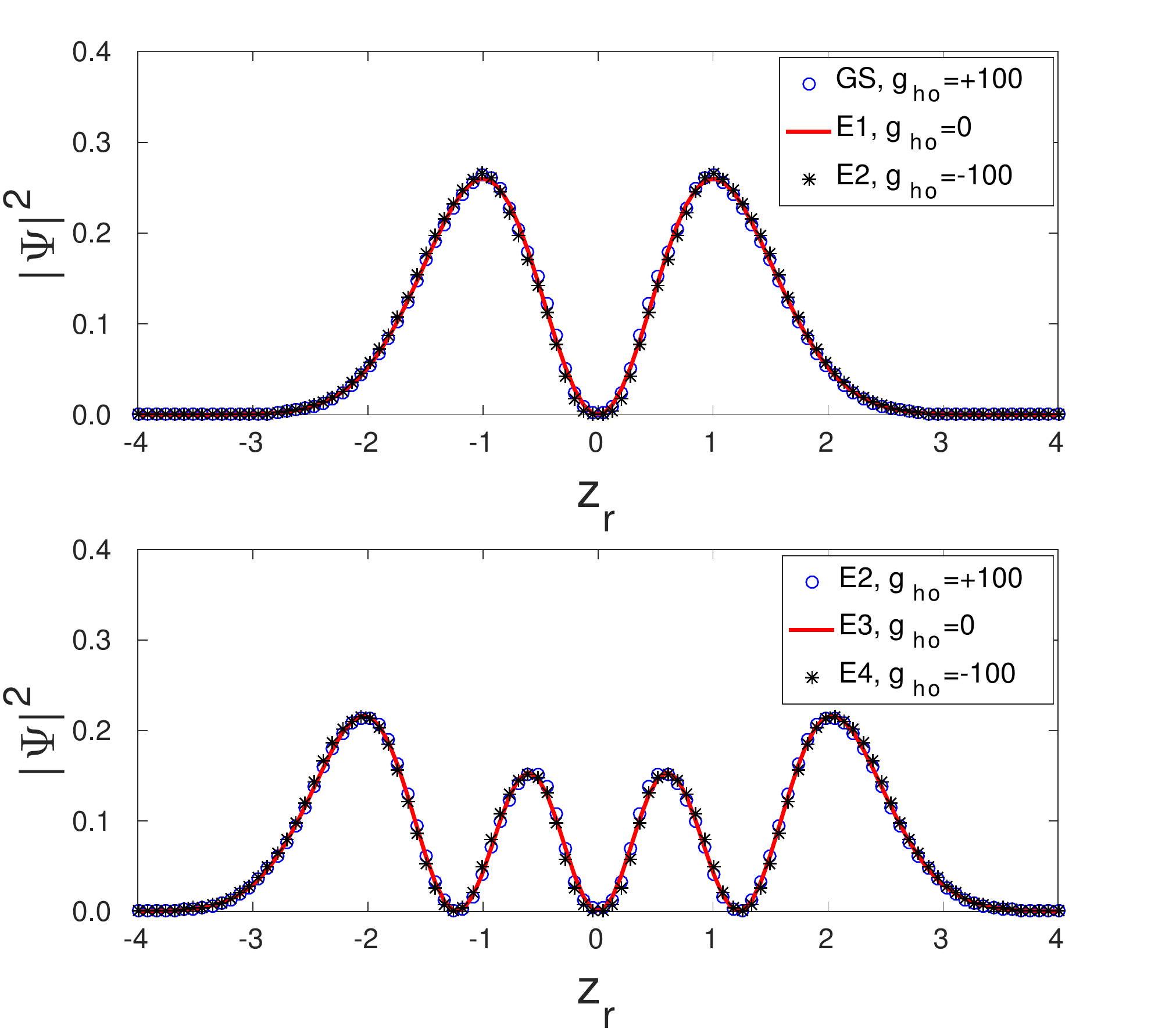}
\caption{(top panel) Comparisons of the probabilities for the ground state relative wave function vs the dimensionless relative coordinate,
$z_r \equiv x_r \sqrt{\mu \omega/\hbar}$ for  $g_{\rm ho} = +100$ (blue circles)
with the non-interacting fermion case (red curve), and with the excited state for large attractive interaction strength 
($g_{\rm ho} = -100$ (black asterisks). The agreement with the non-interacting fermion case demonstrates the `fermionization'
that takes place, both for strong repulsive and for strong attractive interaction strength. In the bottom panel the same comparisons
are made for the next excited state.}
\label{figure9}
\end{figure}

One can follow the progression of the two-particle wave function as the particle-particle interaction varies. The ground state clearly
has bosonic character. As $g_{\rm ho} \rightarrow -\infty$, the two particles remain close together; this is illustrated by the top right
($g_{\rm ho} = -100$) frame in Fig.~\ref{Fig: Wavefun comp}, where the positions of the two particles are clearly strongly correlated 
($x_1$ large means $x_2$ is large as well), or the solid (blue) curve in the top right frame of 
Fig.~\ref{Fig: Wavefun comp_relative}, where the relative wave function is peaked at $x_r = 0$. In contrast, the top left panel
in Fig.~\ref{Fig: Wavefun comp} shows that when the interaction potential is strongly repulsive ($g_{\rm ho} = +100$), the 
two particles avoid one another as best they can, within the confines of the harmonic potential.  This view is reinforced
in Fig.~\ref{Fig: Wavefun comp_relative}.

The lower left panel in both figures is the fermion case (for any interaction strength---here we used $g_{\rm ho} = 0$), which
illustrates the ``fermionization'' taking place in the top left panel, since the two appear to be identical.  We also show the ground state for $g_{\rm ho} = 0$ (a bosonic state)
in the bottom right panel for reference.

One can also examine the 2nd excited state (see Fig.~\ref{Fig: Erel AGS}). Plots of the relevant wave functions are shown in 
Fig.~\ref{figure9}(a), first for $g_{ho} = -100$, where the quantitative agreement with the two left panels in
Fig.~\ref{Fig: Wavefun comp_relative} is apparent. Also shown in Fig.~\ref{figure9}(b) is the probability for the 2nd excited state, 
for $g_{\rm ho} = 100$,
compared with the non-interacting fermion state, with energy just above it, and with the 4th excited state for $g_{\rm ho} = -100$
(again see Fig.~\ref{Fig: Erel AGS}). All three of these probabilities look identical. Clearly ``fermionization'' occurs in the excited states as well, and for large negative values of the coupling strength as well as for large positive values.

\section{Three or more Particles with Interactions \label{secfour}}

Beyond two particles, the methodology of the solution changes; hence we summarize the key elements involved. The most 
straightforward approach is again to view the many-particle wave function in terms of product states of the single particle wave functions.
Matrix elements involving the kinetic energy and the trapping potential are as simple as with two particles; the
third particle (and all other particles beyond two) acts as a ``spectator'' and is unaffected by the interaction. Writing Eq.~(\ref{eq:gen_ham})
explicitly for 3 particles, we have
\begin{equation}
\begin{split}
\hat{H}&=\hat{H}_{ho}(x_1)+\hat{H}_{ho}(x_2)+\hat{H}_{ho}(x_3)\\
&+\hat{V}_{int}(x_2-x_1)+\hat{V}_{int}(x_3-x_1)+\hat{V}_{int}(x_3-x_2)
\end{split}
\label{Eq: 3particle Hamiltonian}
\end{equation}
where $\hat{H}_\textrm{ho}(x_i)$ includes both the kinetic energy and the harmonic oscillator confining potential of the $i$th particle. 

\begin{figure}[t]
\includegraphics[width=0.48\columnwidth]{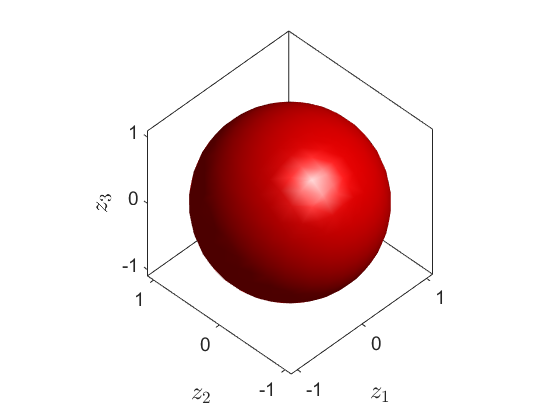}
\includegraphics[width=0.48\columnwidth]{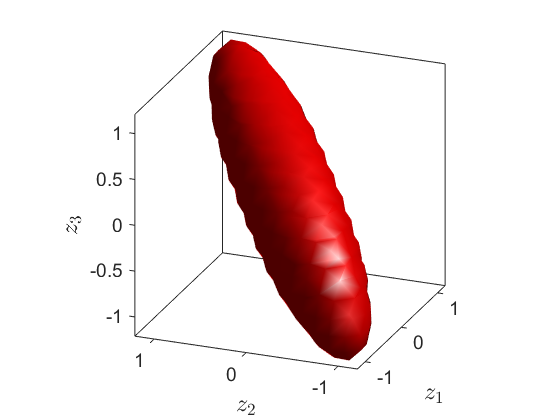}
\includegraphics[width=0.48\columnwidth]{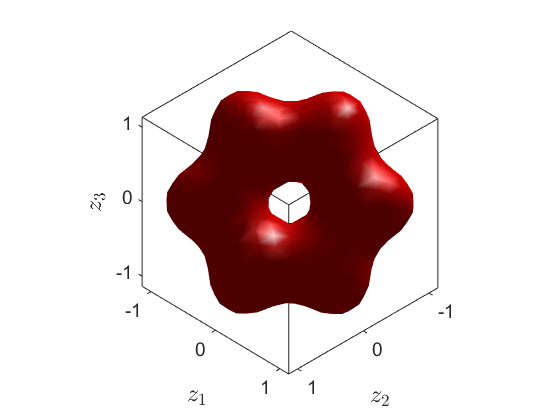}
\includegraphics[width=0.48\columnwidth]{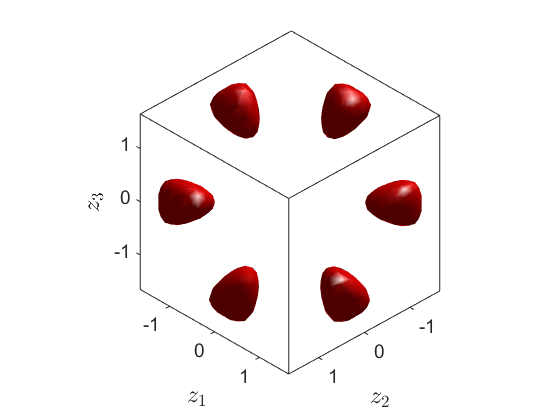}
\caption{Isosurfaces for the three-particle wave function for (a) $g_{ho} = 0$, (b) $g_{ho} = -10$, (c) $g_{ho} = +1$, and (d) 
$g_{ho} = +10$. The figures are all plotted for $|\psi(z_1,z_2,z_3)|^2 = 0.3$. In the second frame it is apparent that a strong
attractive interaction correlates the positions of the three particles so they try to stay on top of one another as much as possible.
In contrast the bottom right frame indicates that nonzero probability tends to occur when the particles are separated from one another,
with a peak at $z_1 = 1$, $z_2 = -1$, and $z_3 = 0$, plus all possible permutations of these.}
\label{figure10}
\end{figure}

To proceed further, one can specify the nature of the particles: distinguishable, fermion, or boson.  We will proceed just with the distinguishable case, but for the sake of completeness, we specify how the states would be enumerated in each case.
Using Dirac bra-ket notation, we specify a 3 particle state as
\begin{equation}
|n,m,l\rangle \equiv |\phi_n(x_1)\phi_m(x_2)\phi_{\ell}(x_3)\rangle,
\label{3particle_basis}
\end{equation}
where $\phi_n(x_1)$ is the single particle harmonic oscillator state as written in Eq.~(\ref{Eq: one particle HO basis}),
and again we focus on the distinguishable case. Basis states are denoted by
\begin{equation}
|\psi_{n,m,\ell}\rangle=|n,m,\ell\rangle 
\label{disting}
\end{equation}
for $n = 0,1,2,3,\ldots\,$, $m = 0,1,2,3,\ldots\,$, $\ell = 0,1,2,3,\ldots\,$. For purposes of enumeration, a sensible ordering of the states 
would be according to their non-interacting energy total, proportional to the sum of the quantum numbers. Hence we 
would require basis states with quantum numbers
\begin{eqnarray}
(n,m,\ell) &=& (0,0,0),\nonumber \\
& & (0,0,1), (0,1,0), (1,0,0), \nonumber \\
& & (0,1,1), (1,0,1), (1,1,0), (2,0,0), (0,2,0), (0,0,2),\nonumber \\
& & (1,1,1), \ldots\,,
\label{qu_nos_dist}
\end{eqnarray}
and it is clear that each row contains states that are degenerate in total non-interacting energy. The boson and
fermion cases are listed in Appendix A.

Naturally we have to truncate, and after some experimentation we have chosen to truncate according to the manner just presented,
i.e., using states up to some maximum sum of the three quantum numbers, 
$n_{\rm tot}$. So, for example, in the list (\ref{qu_nos_dist}) the last line has $n_{\rm tot} = 3$, whereas in the list in Appendix A,
Eq.~(\ref{qu_nos_ferm_app}), the last line has $n_{\rm tot} = 7$. For the distinguishable case it is easy to see that this implies $N_{\rm max}$
basis states, with $N_{\rm max} = (n_{\rm tot} + 1)  (n_{\rm tot} + 2) (n_{\rm tot} + 3)/6$. Thus, even for a modest $n_{\rm tot} = 30$
one has to diagonalize a $5456 \times 5456$ matrix.

The matrix elements are readily calculated, as in the two particle case. Using the shorthand $n \equiv (n_1,n_2, n_3)$ and 
$m \equiv (m_1,m_2, m_3)$, in general we need
\begin{equation}
h_{n,m} \equiv {h_{\rm ho}}_{n,m} + {v_{\rm int}}_{n,m},
\label{mat_ele_def}
\end{equation}
where
\begin{eqnarray}
{h_{\rm ho}}_{n,m}  &=& \langle n_1,n_2,n_3 | \hat{h}_{\rm ho} | m_1, m_2, m_3 \rangle, \\
{v_{\rm int}}_{n,m} &=& \langle n_1,n_2,n_3 | \hat{v}_{\rm int} | m_1, m_2, m_3 \rangle,
\end{eqnarray}
and $\hat{h}_{\rm ho} \equiv \hat{H}_{\rm ho}/(\hbar \omega)$ and $\hat{v}_{\rm int} \equiv \hat{V}_{\rm int}/(\hbar \omega)$ refer to the dimensionless
versions of the first three and second three terms, respectively, of Eq.~(\ref{Eq: 3particle Hamiltonian}).
The first of these is straightforward,
\begin{equation}
{h_{\rm ho}}_{n,m} = \left(n_1+n_2+n_3+\frac{3}{2}\right)\delta_{n_1,m_1}\delta_{n_2,m_2}\delta_{n_3,m_3},
\label{ho_mat_ele}
\end{equation}
while the second can be written in terms of the integral from Eq.~(\ref{vint}), as expressed in Eq.~(\ref{Eq:hermite simplify}).  Thus,
defining (see Eq.~(\ref{c_defn}) for the definition of the constant $c$)
\begin{equation}
{\tilde{v}}_{(n_1,n_2),(m_1,m_2)} \equiv g_{\rm ho} c I(n_1,n_2,m_1,m_2),
\label{vdefn}
\end{equation}
we have, for the three particle case,
\begin{eqnarray}
v_{{\rm int}_{n,m}} &=& v_{{\rm int}_{(n_1,n_2),(m_1,m_2)}}\delta_{n_3,m_3} \nonumber \\
&+& v_{{\rm int}_{(n_3,n_2),(m_3,m_2)}}\delta_{n_1,m_1} \nonumber \\
&+& v_{{\rm int}_{(n_1,n_3),(m_1,m_3)}}\delta_{n_2,m_2}.
\end{eqnarray}

\begin{figure}[t]
\includegraphics[width=1\columnwidth]{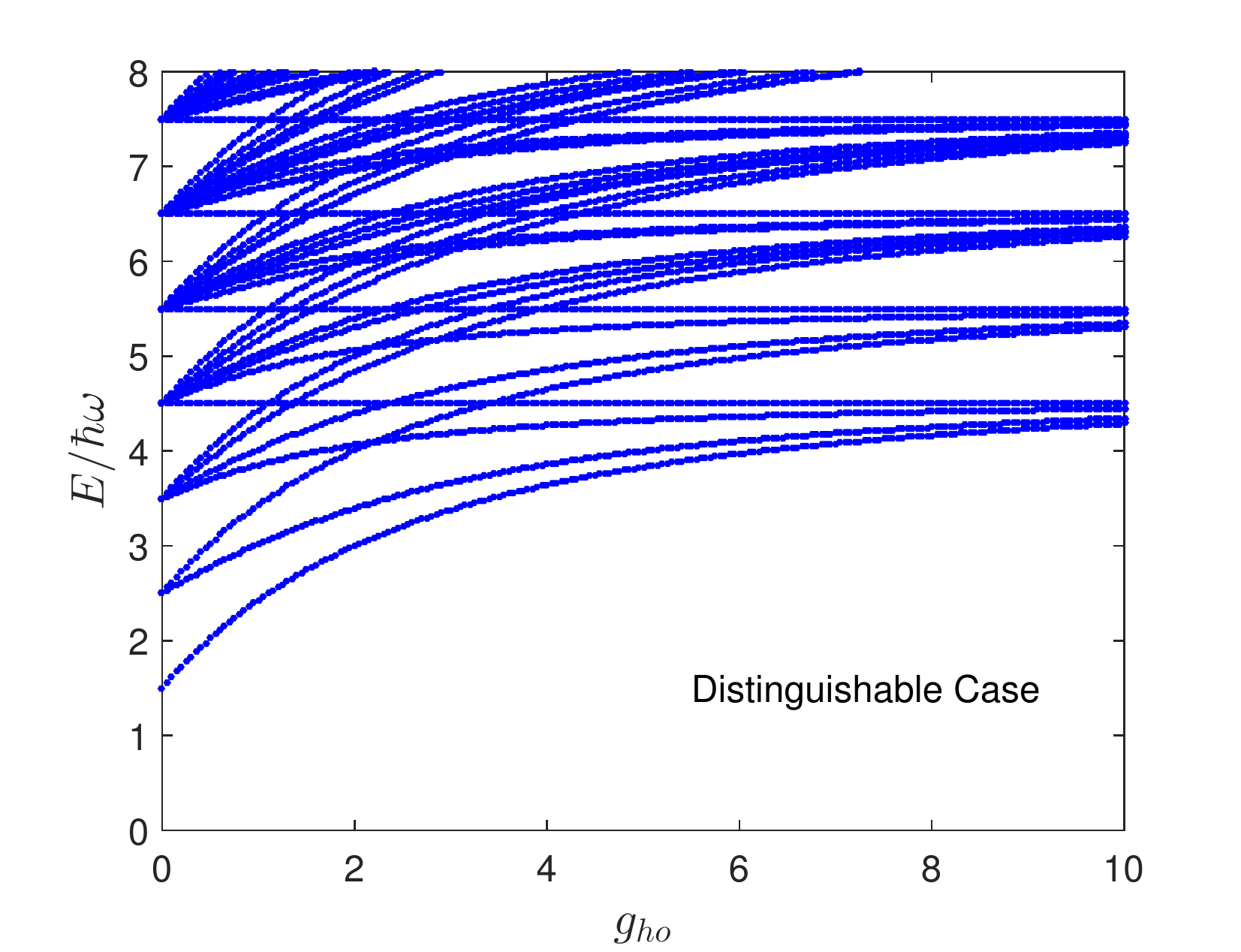}
\caption{Energy levels for three particles in a parabolic trap, shown for repulsive interactions only. Fermionization is evident
for the strongly repulsive case, as seen on the far right of this figure. Similar fermionization occurs for strongly attractive
interactions; these are not shown because the density of levels in the attractive regime is too high.
We used $N_{\rm max} = 5455$. }
\label{figure11}
\end{figure}

Note that for three or more particles, one can again separate out the center-of-mass motion, and focus on the remaining
degrees of freedom. We do not pursue this separation procedure here.\cite{liu10}

Figure~\ref{figure10} illustrates isosurfaces of the wave functions for the three particle case. We have plotted surfaces of constant probability
as a function of the three dimensionless coordinates, $z_1$, $z_2$, and $z_3$, for various values of the dimensionless coupling
constant $g_{\rm ho}$. For very large attractive coupling ($g_{\rm ho} = -10$), the three particles are essentially on top of one another,
while for very large repulsive coupling they clearly avoid one another. The six-fold symmetry in this figure reflects the
fact that the state is bosonic, and hence one requires a wave function that is symmetric in the three coordinates.
Figure~\ref{figure11} shows their energy levels. Clearly separation into bosonic and fermionic states continues to occur, though many of these levels correspond
to states that are distinguishable only, i.e., they are neither bosonic nor fermionic.

\section{Summary}

We have outlined a straightforward methodology to determine the energy eigenstates and eigenvalues for two and three interacting
particles confined in a trapping potential, focusing mainly on harmonic oscillator trap. For students who have been exposed to
numerical matrix mechanics,\cite{marsiglio09} including interactions in this way represents a minor extra step. The more difficult part is
to become familiar with a many-body wave function.  By studying two or three particles, and by using a variety of analytical
and numerical procedures, we hope to have made this next
step easier for the novice. We also demonstrated the concept of ``fermionization,'' which is a first glimpse at the impact of the
indistinguishability of identical particles.  Fermionization occurs for both strongly repulsive and strongly attractive interactions, and
occurs for the excited states as well as the ground state in this problem.

\begin{acknowledgments}
We would like to acknowledge preliminary work performed for this problem by Dylan Grandmont, Collin Tittle, and Noel Hoffer.
This work was supported in part by the Natural Sciences and Engineering
Research Council of Canada (NSERC). In addition, this work was made possible in part by an NSERC USRA (Undergraduate
Student Research Award) to MengXing Na,
and originated in work originally funded by a University of Alberta Teaching and Learning Enhancement Fund (TLEF) grant.
\end{acknowledgments}

\appendix

\section{Many-body wave functions and matrix elements}

The many-body wave function is in general a complicated function of many variables. Very often
significant advances in physics occur when someone manages to come up with a creative representation
of such a wave function, which serves to capture important correlations amongst the particles. In the absence
of such flashes of insight however, the most straightforward way to proceed is with a basis set consisting of
product states of the single particle basis states. Rather than give a general description as found in many-body
textbooks, we will use explicitly the two-particle and three-particle cases as examples, as used in the main body of this paper.

For two particles, we have in principle three cases, distinguishable particles,
\begin{equation}
\psi_{n_1,n_2}=\phi_{n_1}(x_1)\phi_{n_2}(x_2), \phantom{aaa} ({\text{distinguishable}})
\label{dist_app}
\end{equation}
fermions,
\begin{eqnarray}
\psi_{n_1,n_2}=&&\frac{1}{\sqrt{2}}\left[\phi_{n_1}(x_1)\phi_{n_2}(x_2)-\phi_{n_2}(x_1)\phi_{n_1}(x_2)\right],  \nonumber \\
&&\phantom{aaaaaaaaaaaaaaaaaa}(\text{fermions})
\label{fermi_app}
\end{eqnarray}
and bosons,
\begin{eqnarray}
\psi_{n_1,n_2}&&=\begin{cases}
		\phi_{n_1}(x_1)\phi_{n_2}(x_2) & n_1=n_2 \nonumber \\
		\frac{1}{\sqrt{2}}\left[\phi_{n_1}(x_1)\phi_{n_2}(x_2)+\phi_{n_2}(x_1)\phi_{n_1}(x_2)\right] & n_1\neq n_2 
		\nonumber \\
		 \phantom{aaaaaaaaaaaaaaaaaaaaa}(\text{bosons}) \\
		\end{cases}\\
\label{bosons_app}
\end{eqnarray}
where $n_1=1,2,3,...\quad$ and $n_2=1,2,3,...\quad$ for the distinguishable case, while $n_1 > n_2$ only, 
for both the fermion case and for the second line in Eq.~(\ref{bosons_app}) of the boson case.
As used in Section (II.B) and Section (III.A) the single particle wave functions are those of Eq.~(\ref{Eq: ISW basis}).
However, starting in Section (III.C) the single particle wave functions are those of Eq.~(\ref{Eq: one particle HO basis}).
For the former, evaluation of the relevant matrix elements using the products of the single particle wave functions,
Eq.~(\ref{Eq: ISW basis}) results in diagonal elements, for the distinguishable case,
\begin{equation}
{h_0}_{n,m} \equiv {H_0}_{n,m}/E_1 =(n_1^2+n_2^2)\delta_{n_1,m_1}\delta_{n_2,m_2} 
\phantom{a} (\text{dist})
\label{matd_app}
\end{equation}
the fermion case,
\begin{equation}
{h_0}_{n,m}=(n_1^2+n_2^2)(\delta_{n_1,m_1}\delta_{n_2,m_2}-\delta_{n_1,m_2}\delta_{n_2,m_1})
 \phantom{a} (\text{ferm})
\label{matf_app}
\end{equation}
and for the boson case,
\begin{equation}
\begin{split}
&{h_0}_{n,m}=\begin{cases}
(n_1^2+n_2^2)(\delta_{n_1,m_1}\delta_{n_2,m_2}+\delta_{n_1,m_2}\delta_{n_2,m_1}) &\\
\ \ \ \ \ \ \ \ \ \ \ \ \ \ \ \ \ \ \ \text{if}\quad n_1\neq n_2 \quad\&\quad m_1\neq m_2 \\
2n_1^2\delta_{n_1,m_1} \ \ \ \ \ \ \ \text{if}\quad n_1= n_2 \quad\&\quad m_1= m_2 \\
0 \ \ \ \ \ \ \ \ \ \ \ \ \ \ \ \ \ \ \text{otherwise.} \ \ \ \ (\text{boson})
\label{matb_app}
\end{cases}
\end{split}
\end{equation}
where $E_1=\pi^2\hbar^2/(2m_0a^2)$, which is the single particle ground state energy of the infinite square well. 
We should mention here that the sizes of the Hilbert spaces vary, depending on the particle statistics. The case of two particles
is very special; the Hilbert space for distinguishable particles happens to equal the sum of the sizes of the fermion and boson
Hilbert spaces, so that one can say that the states are conserved as statistics applicable to indistinguishable particles is introduced.
However, in general, as the number of particles increases, the size of the Hilbert space pertaining to distinguishable particles greatly
exceeds the size of the other two spaces.

With interactions,
we obtain both diagonal and off-diagonal matrix elements,
\begin{equation}
{V_{\rm int}}_{n,m}=\langle\psi_{n_{1},n_{2}}|\hat{V}_{\rm int}|\psi_{m_{1},m_{2}}\rangle.
\label{Dirac interaction matrix elements_app}
\end{equation}
For the contact interaction, Eq.~(\ref{Eq: Dirac-delta interaction potential}), and distinguishable statistics, we obtain 
the result Eq.~(\ref{con_disting}) with the matrix element defined in Eq.~(\ref{dim_pot_mat_ele}). Using that notation, it is
clear that the result for fermions (F) is ${V_{\rm int F}}_{n,m} = 0$. For bosons (B) the result is
\begin{equation}
{V_{\rm int B}}_{n,m} =\begin{cases}
		2V(n_1,n_2;m_1,m_2) \ \text{if}\ n_1\neq n_2 \ \& \ m_1\neq m_2 \\
		V(n_1,n_2;m_1,m_2) \ \ \ \text{if} \ n_1= n_2 \ \& \ m_1= m_2 \\
		 \sqrt{2} V(n_1,n_2;m_1,m_2) \ \ \ \  \text{otherwise.} \\
		\end{cases} \\
\label{con_bosons_app}
\end{equation}
As is apparent for the contact interaction, fermions do {\it not} interact with one another at all, and the matrix elements for bosons
are generally larger than or equal to those for distinguishable particles, since their statistics cause bosons to spend more of their
time in contact with one another.

In the case of harmonic oscillator basis states, as used in Section (III.C) and beyond, the diagonal matrix elements for the
three cases are given by, for the distinguishable case,
\begin{equation}
{h_0}_{n,m}=(n_1+n_2+1)\delta_{n_1,m_1}\delta_{n_2,m_2},  (\text{distinguishable}) \\
\label{diagd_app}
\end{equation}
for the fermion case,
\begin{eqnarray}
{h_0}_{n,m}&=&(n_1+n_2+1)(\delta_{n_1,m_1}\delta_{n_2,m_2}-\delta_{n_1,m_2}\delta_{n_2,m_1}) \nonumber \\
&\phantom{a}&\phantom{aaaaaaaaaaaaaaaaaaaaaa} (\text{fermion})
\label{diagf_app}
\end{eqnarray}
and for the boson case,
\begin{equation}
\begin{split}
&h_{0_{n,m}}=\begin{cases}
		(n_1+n_2+1)(\delta_{n_1,m_1}\delta_{n_2,m_2}+\delta_{n_1,m_2}\delta_{n_2,m_1}) &\\
		\quad \quad \quad \quad \text{if}\quad n_1\neq n_2\quad \&\quad m_1\neq m_2 \\
		(n_1+n_2+1)\delta_{n_1,m_1} &\\
		\quad \quad \quad \quad \text{if}\quad n_1=n_2\quad \&\quad m_1=m_2\\
		0 &\\
		 \quad \quad \quad \quad \text{otherwise.}\ \ \ \ \ \ \ \ \ \ \ {\rm (bosons)}\\
		\end{cases}\\	
\end{split}
\label{diagb_app}
\end{equation}
%
where the dimensionless
matrix elements are defined this time by dividing all energies by $\hbar \omega$, i.e. 
${h_0}_{n,m} \equiv {H_0}_{n,m}/\hbar \omega$ [recall,
following the notation of Eq.~(\ref{non_int_ham}), $n$ is shorthand for $(n_1,n_2)$, etc.].

For three particles, the distinguishable case is given by Eq.~(\ref{disting}), with an initial enumeration provided in
Eq.~(\ref{qu_nos_dist}). 
For bosons, a basis state must be symmetric, so we have,
\begin{equation}
\begin{split}
|\psi_{n,m,\ell}\rangle=\begin{cases}
	\frac{1}{\sqrt{6}}\Big(|n,m,\ell\rangle+|n,\ell,m\rangle+|m,n,\ell\rangle+|m,\ell,n\rangle\\
	\ \ +|\ell,n,m\rangle+|\ell,m,n\rangle \Big) \phantom{aaa} \text{if}\quad n > m > \ell&\\
	\phantom{aa} \\
	\frac{1}{\sqrt{3}}(|n,n,\ell\rangle+|n,\ell,n\rangle+|\ell,n,n\rangle )&\\
	\phantom{aaaaaaaaaaaaaaaaaaaaaa} \text{if}\quad n= m > \ell\\
	\phantom{aa} \\
	|n,n,n\rangle \phantom{aaaaaaaaaaaaaaaa} \text{if} \quad n=m=\ell. \\
\end{cases}\\
\end{split}
\label{bos_app}
\end{equation}
Thus, an enumeration of the basis states proceeds as
\begin{eqnarray}
(n,m,\ell) =& & (0,0,0),\nonumber \\
& & (1,0,0), \nonumber \\
& & (2,0,0), (1,1,0), \nonumber \\
& & (3,0,0), (2,1,0), (1,1,1)\nonumber \\
& & (4,0,0), (3,1,0), (2,2,0), ... \text{bosons} 
\label{qu_nos_bos_app}
\end{eqnarray}
with each row degenerate in total non-interacting energy.

Finally, for fermions, a basis state must be antisymmetric, so that basis states are given by the usual Slater determinant,
Finally, for fermions, a basis state must be antisymmetric, so that basis states are given by the usual Slater determinant,
\begin{eqnarray}
|\psi_{n,m,\ell}\rangle=\frac{1}{\sqrt{6}}&& \bigg(|n,m,\ell\rangle-|n,\ell,m\rangle-|m,n,\ell\rangle+|m,\ell,n\rangle \nonumber \\
&&+|\ell,n,m\rangle-|\ell,m,n\rangle \bigg),
\label{fermion_app}
\end{eqnarray}
now for $n > m > \ell$. Hence an enumeration of the basis states proceeds as
\begin{eqnarray}
(n,m,\ell) =& & (2,1,0),  \phantom{aaaaaaaaaaaaaaaaaaaa} \text{fermions} \nonumber \\
& & (3,1,0), \nonumber \\
& & (3,2,0), (4,1,0), \nonumber \\
& & (3,2,1), (4,2,0), (5,1,0)\nonumber \\
& & (4,2,1), (4,3,0), (5,2,0), (6,1,0), ....
\label{qu_nos_ferm_app}
\end{eqnarray}

\section{Dirac-delta Wang trick}

The Dirac delta interaction term is given by $\hat{V}_{\rm int}=g*\delta(x_{1}-x_{2})$, so that the required matrix
element (for two particles --- see Eq.~(\ref{Eq: integral Dirac-delta matrix element})) is
 \begin{equation}
 {v_{\rm int}}_{n,m}=cg_{\rm ho}\int_{-\infty}^{\infty}H_{n_{2}}(z)H_{m_{2}}(z)H_{n_1}(z)H_{m_1}(z)e^{-2z^{2}}dz
 \label{vint_app}
 \end{equation}
Our goal is to solve this integral. We want to take advantage of orthonormality, i.e.
\begin{equation}
\int_{-\infty}^{\infty}e^{-z^2}H_j(z)H_k(z)dz=2^j j!\sqrt{\pi}\delta_{jk},
\end{equation}
and we do this by re-expressing products of Hermite polynomials in $z$ as new Hermite polynomials in $\sqrt{2}z$,
\begin{equation}
H_j(z)H_k(z)=\sum_{r=0}^{j+k}a_r(j,k)H_r(\sqrt{2}z).
\end{equation}
We assume here that $i$ and $j$ have been chosen amongst the 4 possible quantum numbers in Eq.~(\ref{vint_app})
so that their sum is the lowest possible of the 6 combinations.
If we realize that the Hermite polynomials can always be expressed in this way, then:
\begin{equation}
\begin{split}
I=&\int_{-\infty}^{\infty}H_{j}(z)H_{k}(z)H_{p}(z)H_{q}(z)e^{-2z^{2}}dz\\
=&\sum_{n=0}^{j+k}\sum_{m=0}^{p+q}a_n(j,k) a_m(p,q) \times \\
&\quad \int_{-\infty}^{\infty}H_n(\sqrt{2}z)H_m(\sqrt{2}z) e^{-(\sqrt{2}z)^2}dz\\
\text{} =&\sum_{n=0}^{j+k}\sum_{m=0}^{p+q}a_n(j,k) a_m(p,q) 2^m m!\sqrt{\frac{\pi}{2}} \delta_{n,m}\\
\text{} =&\sum_{m=0}^{\ell_{\rm max}}a_m(j,k) b_m(p,q)2^m m!\sqrt{\frac{\pi}{2}},
\end{split}
\label{smax}
\end{equation}
where $\ell_{\rm max}$ in this case is simply $j+k$, and $m$ in the last line is over even (odd) numbers only if $j+k$ is even (odd). Note that
if $j+k$ is even (odd) then $p+q$ is also even (odd) for all nonzero integrals. Furthermore,
note that $p$ and $q$ need not necessarily be quantum numbers inside the same quantum state. Given $2$ states and $4$ quantum numbers (say, $(0, 16)$ and $(1,25)$), we can pick the smallest combination to limit the number of sums we have to do --- $p=0$ and $q=1$ --- so that $\ell_{\rm max} = 1$, and there is only one term in the final sum of Eq.~(\ref{smax}).

The next step is to solve for $a_m(j,k)$.
To do so we take a product of the generating functions:
\begin{equation}
e^{2tx-t^2}=\sum_{n=0}^{\infty}H_n(x)\frac{t^n}{n!}\qquad e^{2sx-s^2}=\sum_{m=0}^{\infty}H_m(x)\frac{s^m}{m!},
\label{gen_her}
\end{equation}
which gives,
\begin{equation}
e^{2(t+s)x-(t^2+s^2)}=\sum_{j=0}^{\infty}\sum_{k=0}^{\infty}H_j(x)H_k(x)\frac{t^j s^k}{j!k!}.
\label{rhs}
\end{equation}
Then we rearrange the left-hand-side (LHS) into two different exponentials, and treat the first as a generating function for Hermite
polynomials, i.e. use the expansion, Eq.~(\ref{gen_her}), and simply Taylor-expand the second. We obtain
\begin{equation}
\begin{split}
{\rm LHS}=&e^{2\frac{t+s}{\sqrt{2}}\sqrt{2}x-(\frac{t+s}{\sqrt{2}})^2}e^{-\frac{1}{2}(t-s)^2}\\
\text{} =&\sum_{\ell=0}^{\infty}H_\ell(\sqrt{2}x)\frac{(t+s)^\ell}{2^{{\ell/2}}\ell!}\sum_{r=0}^{\infty}(-1)^r(t-s)^{2r}\frac{1}{2^r r!}\\
\text{} =&\sum_{\ell=0}^{\infty}H_\ell(\sqrt{2}x)\frac{1}{2^{{\ell/2}}\ell!}\sum_{r=0}^{\infty}\frac{(-1)^r}{2^r r!}\\
&\sum_{u=0}^{\ell}\sum_{v=0}^{2r}C_v^{2r}C_u^\ell(-1)^vs^{v+u}t^{2r+\ell-u-v}.
\end{split}
\label{lhs}
\end{equation}
where $C_u^\ell$ are the binomial coefficients:
\begin{equation}
C_u^\ell \equiv {\ell ! \over (\ell - u)! u !}.
\label{comb_app}
\end{equation}

Now with the LHS represented by Eq.~(\ref{lhs}) and the RHS represented by Eq.~(\ref{rhs}), it must be true that the coefficients
of $t^js^k$ must be the same. This immediately implies
\begin{eqnarray}
v+u&=&k  \nonumber \\
\text{} 2r + \ell -u-v &=& j.
\label{first_two}
\end{eqnarray}
The first can be used to eliminate $v = k-u$ on the LHS, while the second, in conjunction with the first, is to be used
to eliminate $r = (j+k-\ell)/2$. This can be immediately substituted into the last equation on the LHS, but it is best to
first note several other consequences on the remaining two sums over $u$ and $\ell$. First, because $r \ge 0$, the
replacement $r = (j+k-\ell)/2$ implies that $\ell \le j+k$. Furthermore, since $2r$ is obviously always even, then if $j+k$ is
even, so too must $\ell$ be, while if $j+k$ is odd, then $\ell$ will be odd. This means that the summation over $\ell$ is
terminated at $j+k$, and starts at zero or one, depending on whether $j+k$ is even or odd, respectively. The fact that
$\ell$ has the same parity as $j+k$ will be indicated by $\ell \sim j + k$ in the summations.

Furthermore there are restrictions on the summation over $u$. The last line of Eq.~(\ref{lhs}) indicates that the maximum value
of $u$ is $\ell$. However, since originally $v \ge 0$, then the first of Eq.~(\ref{first_two}) also implies $u \le k$. Therefore
$u \le min(k,\ell)$. The starting value for the remaining $u$ summation can also vary. Since $v \le 2r$ then,
using the first and second lines of Eq.~(\ref{first_two}) for $v$ and $2r$ respectively, we obtain $k-u \le j+k-\ell$ which implies $u \ge \ell -j$.
Since this can be both positive or negative, then $u_{min} = max(\ell -j,0)$.

Inserting all these conditions into the last two lines of Eq.~(\ref{lhs}), and equating the coefficients of $t^js^k$, we obtain
\begin{equation}
\begin{split}
H_j(x)H_k(x)&=\frac{j!k!}{2^{\frac{j+k}{2}}}\sum_{\ell=0,\ell\sim j+k}^{j+k}H_\ell(\sqrt{2}x)\frac{(-1)^{\frac{j+k-\ell}{2}}}{(\frac{j+k-\ell}{2})!\ell!}\\
&\sum_{u=max(\ell-j,0)}^{min(k,\ell)}C_{k-u}^{j+k-\ell}C_u^\ell(-1)^{k-u}.
\end{split}
\end{equation}
The coefficient $a_\ell$ is then
\begin{equation}
a_\ell(j,k)=\frac{j!k!}{2^{\frac{j+k}{2}}}\frac{(-1)^{\frac{j+k-\ell}{2}}}{(\frac{j+k-\ell}{2})!\ell!}\sum_{u=max(\ell-j,0)}^{min(k,\ell)}C_{k-u}^{j+k-\ell}C_u^\ell(-1)^{k-u},
\end{equation}
and our integral is given very simply as
\begin{equation}
\begin{split}
I(j,k,p,q)&=\sum_{\ell=0}^{\ell_{\rm max}}a_\ell(j,k) a_\ell(p,q)2^\ell \ell!\sqrt{\pi \over 2}.\\
\end{split}
\end{equation}
with $\ell_{\rm max}$ defined as below Eq.~(\ref{smax}).


\begin{thebibliography}{99}
\bibitem{remark1} It is noteworthy, however, that experimentalists are forever attempting to `flatten' the confining potential, in an
attempt to eliminate inhomogeneities in the particle gas, so comparisons can be made with the theory for the homogenous gas,
about which more properties can be understood.
\bibitem{remark2} Beware, however, that different conventions are followed by physicists and mathematicians. We use the physicists'
convention, where $H_n(z)$ has a coefficient of $z^n$ equal to $2^n$, i.e. $H_0(z) = 1$, $H_1(z) = 2z$, $H_2(z) = 4z^2 -2$, etc.
\end{thebibliography}
\end{document}